\documentclass[pr,twocolumn,preprintnumbers,nofootinbib,showpacs,amsmath,amssymb,superscriptaddress]{revtex4-1}
\usepackage{psfrag,graphicx}
\usepackage{bigstrut,bigdelim,multirow}
\usepackage[
      colorlinks=true,
      linkcolor=blue,
      urlcolor=blue,
      filecolor=black,
      citecolor=blue,
            ]{hyperref}
\usepackage{color}
\usepackage{amsfonts}
\usepackage{amssymb}
\usepackage{epstopdf}
\usepackage{dsfont}
\usepackage{epsfig}

\begin{document}
\title{Multiple critical gravitational collapse of charged scalar with reflecting wall}
\author{Rong-Gen Cai}
\email{cairg@itp.ac.cn}
\affiliation{Institute of Theoretical Physics,
Chinese Academy of Sciences, Beijing 100190, China}
\affiliation{School of Physics, University of Chinese Academy of Sciences, YuQuan Road 19A, Beijing 100049, China}

\author{Run-Qiu Yang}
\email{aqiu@itp.ac.cn}
\affiliation{Institute of Theoretical Physics,
 Chinese Academy of Sciences, Beijing 100190, China}

%\pacs{PACS}
%\keywords{keywords}
%\preprint{preprint}
%%%%%%%%%%%%%%%%%%%%%%%%%%%%%%%%%%%%%%
\begin{abstract}
%%%%%%%%%%%%%%%%%%%%%%%%%%%%%%%%%%%%%%
In this paper, we present the results on  the gravitational collapse of charged  massless scalar field  in asymptotically flat spacetime with a perfectly reflecting wall. Differing from  previous works, we study the system in the double null coordinates, by which we could simulate the system until the black hole forms with higher precision  but  less performance time. We investigate the influence of charge on the black hole formation and the scaling behavior near the critical collapses. The gapless and gapped critical behaviors for black hole mass and charge are studied numerically. We find that  they satisfy the scaling laws for critical gravitational collapse but the gapped critical behavior is different from its AdS counterpart.
%%%%%%%%%%%%%%%%%%%%%%%%%%%%%%%%%%%%%%
\end{abstract}
%%%%%%%%%%%%%%%%%%%%%%%%%%%%%%%%%%%%%%
\maketitle
%\tableofcontents

%%%%%%%%%%%%%%%%%%%%%%%%%%%%%%%%%%%%%%
\noindent

\section{Introduction}
\label{Introd}
Since  the pioneering work by Choptuik \cite{Choptuik}, the critical phenomena in gravitational collapse have attracted a lot of attention and are still an active topic in numerical relativity (see Ref. \cite{Gundlach:2007gc} for recent reviews). In Choptuik's original work, he investigated the gravitational collapse of a massless scalar field in spherically symmetric asymptotically flat space-times with one-parameter family of initial data parameterized by $\epsilon$. It was found that such a system exhibits critical phenomena with discrete self-similarity. In the vicinity of the critical solution, the mass contained in the initial apparent horizon obeys following scaling relationship,
\begin{equation}\label{masscaling1}
M_h\propto\left\{
\begin{split}
&(\epsilon-\epsilon^*)^\beta, ~~\epsilon\rightarrow \epsilon^{*+},\\
&0,~~~~\epsilon<\epsilon^*.
\end{split}
\right.
\end{equation}
with the critical solution at $\epsilon=\epsilon^*$ and $\beta\approx0.37$. The scaling exponent $\beta$ is universal, i.e., the same for different family of initial configurations. But it has been also found that  $\beta$ has different values for same special initial function families  \cite{Patrick,Zhang:2014dfa} and is also model dependent~\cite{Maison:1995cc}. Since then, the critical gravitational collapse for different matter fields in open space-time (asymptotic flat or de Sitter (dS) space-time) has been studied extensively both by numerical simulations and theoretical analysis. The critical scaling of the mass can now be  understood in terms of general relativity as a dynamical system, and of Lyapunov's exponents \cite{Gundlach:1995kd,Koike:1995jm}, at least in the spherically symmetric case.

However, the gravitational collapse in closed system is very different. The first one of such a system is the gravitational collapse in asymptotically anti-de Sitter (AdS) space-time. With the abundant of investigations in the holographic duality provided by the anti-de Sitter/Conformal Field Theory (AdS/CFT) correspondence \cite{Maldacena:1997re,Witten:1998qj}, the dynamics of critical collapse and the stability of AdS have been studied in details in the last years. One of important results in the numerical computations for gravitational collapse in asymptotically AdS$_{d+1}$ ($d\geq3$) is the discovery of weakly turbulent instability~\cite{Bizon:2011gg}.   Different from  the asymptotically  flat or dS space-time cases, there are infinite critical solutions for given one-parameter family of initial configuration and the black hole can form with an  infinitesimal energy after enough time. Another closed system was studied by Maliborski in Ref.~\cite{Maliborski:2012gx}, where he studied a spherically symmetric self-gravitating massless scalar field enclosed inside  a perfectly reflecting wall in a flat
spacetime. Numerical evidence is given that arbitrarily small generic initial data evolve into a black hole. This result shows that turbulent dynamics and infinite critical collapses can occur in a very wide system when dissipation of energy by dispersion is absent rather than specific to asymptotically AdS spacetime. The turbulent behavior is not an exclusive domain of asymptotically AdS spacetimes but a typical feature of ``confined"  Einstein's gravity with reflecting boundary conditions and the only role of cosmological constant $\Lambda$ is to generate the timelike boundary at spatial infinity.

The energy transformation and reasons of weakly turbulent behavior have been studied for several years. However, the critical behavior for the multiple critical solutions is still a mysterious topic. The relationship between these infinite critical solutions and how the critical solutions distribute in a given one-parameter initial data family are still open questions and lack of enough study both in numerical computation and analytic analysis.  The recent work in Ref. \cite{Olivan:2015fmy} found that  there is a mass gap when $\epsilon\rightarrow \epsilon_{n}^-$ in the asymptotically AdS spacetime. For the asymptotic AdS case, the Choptuik's scaling relationship can be modified as,
\begin{equation}\label{masscaling2}
\left\{
\begin{split}
&M_h\propto(\epsilon-\epsilon^*)^{\beta^+}, ~~\epsilon\rightarrow \epsilon_n^{+},\\
&M_h-M_{ng}\propto(\epsilon-\epsilon^*)^{\beta^-},~~\epsilon\rightarrow \epsilon_n^-.
\end{split}
\right.
\end{equation}
Here $\epsilon_n$ is the n-th critical value of $\epsilon$, $M_{ng}$ is the n-th mass gap at $\epsilon\rightarrow \epsilon_n^-$,  $\beta^+\approx0.37$, while $ \beta^-\approx0.70$. The new critical exponent $\beta^-$ is found universal. This result shows that the critical gravitational collapses in closed systems are much rich than those in open systems. Different from the case $\epsilon\rightarrow \epsilon_n^+$ where critical phenomenon is determined by local geometry, such gapped critical phenomena should have relationship with overall geometry contained in the initial apparent horizon. So one can expect that the critical exponents in asymptotically AdS spacetimes and flat spacetimes with perfectly reflecting mirror may be different.

In studying such problems, one  of  main difficulties is the challenge of numerical computation. If one is going to study the critical behavior, the numerical solver must have ability to cover the case that an apparent horizon appears in very close to the origin point of the coordinates. This needs the spatial discretization is small enough near the position where the apparent horizon appears. In the uniform spatial grid method (see, for example, in Refs.~\cite{Bizon:2011gg,Maliborski:2012gx}), it at least needs about $10^5$ grids and will spend time from days to weeks in supercomputer to finish numerical simulations for every initial data. The other method that may be suitable for such problems is the adaptive refined algorithm. However, this method is rather complicated and has not been applied into the multiple critical collapse so far.

In this paper, we will propose a new algorithm to solve the gravitational collapse in a closed system and study the influence of charge on the multiple critical phenomena in gravitational collapse. This method was inspired by  Refs. \cite{Gundlach,Garfinkle,Hod:1996ar}. We solve the gravitational collapse in the double-null coordinates with null initial data in a null surface, by which we could simulate the system until the black hole forms with higher precision  but  less performance time. Theoretically, our method has the same precision as that in  Ref. \cite{Olivan:2015fmy} when the black hole nearly forms, but we don't need to make a transformation from the Cauchy-type evolution to characteristic evolution. By this method, we study the gravitational collapse of a charged  massless scalar field  in asymptotically flat spcetime with a perfectly reflecting wall. Our numerical results show that the mass gap also appears in the asymptotic flat space-time with a perfectly reflecting wall but the critical exponent $\beta^-\approx0.36$, which is universal but different from  the one in the asymptotically AdS case.  Except for  the mass gap, the charge contained in the initial apparent horizon also shows the similar gaped scaling behavior. To describe the scaling behavior when $\epsilon\rightarrow \epsilon_n^\pm$, we define  two groups of  critical exponents and compute them numerically. These critical exponents are agreement with the scaling law proposed recently in Ref. \cite{Cai:2015b}.  Here we mention that the gravitational collapse of charged scalar field in asymptotically flat spacetime has been studied  in \cite{Gundlach:1996vv},  and in asymptotically de Sitter space has been recently discussed in \cite{Zhang:2015dwu}.

This paper is organized as follows. In  section \ref{nullcoord}, we will first give out the equations of motion of the system in the double-null coordinates and  introduce the algorithms. In  section \ref{Numsim}, we will explain how to deal with the timelike boundary  conditions in the double-null coordinates, which is main difficulty in such coordinates. Then some tests for the convergency and precision for  our solver will  be shown in this section. In  section \ref{results}, we will study the multiple critical collapse and the scaling relationships numerically. The summary and some discussions will be given in  section \ref{summ}.

\section{Equations of motion in double null coordinates}
\label{nullcoord}
\subsection{Equation of motions}
In 3+1 dimensional asymptotically  flat space-time, the dynamics of Einstein gravity, a charged massless scalar field and  Maxwell field are described by the following action,
\begin{equation}\label{action1}
    S=\int d^4x\sqrt{-g}\left[\frac{\mathcal{R}}{16\pi G}+\mathcal{L}_m\right],
\end{equation}
with
\begin{equation}\label{Lagm}
   \mathcal{L}_m=-\frac12(D_a\phi)(D^a\phi)^\dagger-\frac1{16\pi}F_{ab}F^{ab}.
\end{equation}
Here $\mathcal{R}$ is the scalar curvature, $G$ is the Newton gradational constant, $\phi$ is the complex scalar field. $F_{ab}=(dA)_{ab}$ and $A_a$ is the gauge field. $D_a\phi=\nabla_a\phi+ieA_a\phi$ and $e$ is the charge carried by scalar field. For convenience, we take $G=1$. By variation, we can get the  equations of motion of the system,
\begin{equation}\label{Einscal},
\begin{split}
    \mathcal{R}_{ab}-\frac{\mathcal{R}}2g_{ab}=&8\pi T_{ab},\\
    D_aD^a\phi=&0,\\
   \frac1{4\pi}\nabla^bF_{ab} -ie[\phi(D_a\phi)^\dagger-\phi^\dagger D_a\phi]=&0.
    \end{split}
\end{equation}
Here $\mathcal{R}_{\mu\nu}$ is the Ricci tensor. $T_{ab}$ is the energy momentum tensor of the matter sector,
\begin{equation}\label{energyT}
T_{ab}=(D_{(a}\phi)(D_{b)}\phi)^\dagger+\frac1{4\pi}{F_a}^cF_{bc}+g_{ab}\mathcal{L}_m.
\end{equation}
Our ansatz for metric and gauge field in the double null coordinates can be described by~\cite{Garfinkle,Hod:1996ar},
\begin{equation}\label{globm}
    ds^2=-fr'dudv+r^2d\Omega^2,~~A_a=[A(u,v),0,0,0].
\end{equation}
Here $u$ and $v$ are two null coordinates, $r$ and $f$ are the functions of $u$ and $v$. Here we take $u$ as the null time and $v$ as the null spatial coordinate. The prime means the partial derivative with respect to $v$ and a dot means the partial derivative with respect to $u$.

Now introduce  auxiliary variables $\Phi$ and $\bar{f}$ such as,
\begin{equation}\label{barphhi}
\phi=\frac1{r}\int_{v_0(u)}^v\Phi r'dv,
\end{equation}
and
\begin{equation}\label{barf}
    \dot{r}=-\bar{f}/2.
\end{equation}
Here $v_0(u)$ is defined by equation $r(u,v_0(u))=0$.  By the $v$-component of equations for gauge field in Eqs.~\eqref{Einscal}, the charge contained within the sphere
of radius $r$ at a retarded time $u$ is expressed as,
\begin{equation}\label{charger}
    Q(u,v)=4\pi ie\int_{v_0(u)}^v(\phi^\dagger\Phi-\phi\Phi^\dagger)rr'dv,
\end{equation}
and the gauge potential is,
\begin{equation}\label{guageA}
    A(u,v)=\int_{v_0(u)}^v\frac{Q}{r^2}fr'dv.
\end{equation}
The $u$-component of equations for gauge field is not independent. Take these into the equation for scalar field in \eqref{Einscal}, we obtain a time evolutional equation,
\begin{equation}\label{eqscalar2}
    \dot{\Phi}=\frac1{2r}(\Phi-\phi)[(1-Q^2/r^2)g-\bar{f}]-\frac{ieQg\Phi}{2r}-ie\phi A.
\end{equation}
There are only three nontrivial components of Einstein's equations. By the $uv$ and $vv$ components of Einstein's equations in Eqs.~\eqref{Einscal}, $f$ and $\bar{f}$ can be solved by following integrations,
\begin{equation}\label{ffbar1}
\begin{split}
    f&=F_1(u)\exp\left[4\pi\int_{v_0}^v\frac{|\Phi- \phi|^2r'}{r}dv\right],\\
    \bar{f}&=\frac1{r}\int_{v_0}^v[1-Q^2/r^2]f r'dv+\frac{F_2(u)}{r}.\\
    \end{split}
\end{equation}
Here functions $F_1(v)$ and $F_2(v)$ are defined by boundary conditions. We see that in this system, there are only two time-evolution equations \eqref{barf} and \eqref{eqscalar2}. The solution at a given $v$ depends only on the solution in $\tilde{v}<v$.

The $uu$ component of  Einstein's equations linearly depends on others and can be used to check
the accuracy of numerical computation. By combining the $uu$ and $uv$ components of  Einstein's equations, we have
\begin{equation}\label{chackerror1}
K(u,v)=\partial_u\ln(\bar{f}/f)+\frac{f-\bar{f}}{2r}-\frac{Q^2f}{2r^3}-8\pi rT_{uu}/\bar{f}=0,
\end{equation}
where
\begin{equation}\label{Tuu}
    T_{uu}=|\partial_u\phi|^2+ieA(\phi\partial_u\phi^\dagger-\phi^\dagger\partial_u\phi)+e^2A^2|\phi|^2.
\end{equation}
 Eq.~\eqref{chackerror1} is used to check the precision in the progress of time evolution. The time partial derivatives are computed by 4th order center difference method with previous two steps and later two time steps. The deviation of \eqref{chackerror1} can describe the error of our numerical algorithm.

In spherically symmetric  spacetime, Misner-Sharp mass is a very useful quantity. In the investigation on the gravitational collapse, this mass is a very important because it shows some scaling relation with respect to the parameter of initial family \cite{Choptuik}. The
Misner-Sharp mass function is given by,
\begin{equation}\label{Mass1}
    M(u,v)=\frac{r}2(1-\bar{f}/f+Q^2/r^2).
\end{equation}
Here $r$ is a function of $u,v$ and $Q$ is defined by integration \eqref{charger}.

\subsection{Boundary conditions and gauge fixing}
\label{BCGF}
Now let us consider the boundary conditions. In this paper, we consider the system with a perfectly reflecting mirror at a fixed radius $r=R$ for matter field. There are two boundaries which locate at the curves of $r(u,v)=0$ and $r(u,v)=R$, respectively. In the null coordinates, these two boundaries are both dynamic and evolute with the null time $u$, which bring some difficulties when we perform numerical simulation by finite difference method. We will discuss the details about how to impose boundary conditions in numerical computations in section~\ref{Numsetup}.

First we impose regular conditions for all the functions at the origin point $r=0$. i.e., all the quantities are finite and smooth at $r=0$.  Then we have $F_2(u)=0$ and $F_1(u)$ should be finite. From the expressions in Eqs.~\eqref{barphhi} and \eqref{ffbar1}, this condition implies following Taylor's expansions,
\begin{equation}\label{taylorF}
    \mathcal{F}=\mathcal{F}_0+\mathcal{F}_1r+\mathcal{F}_2r^2+\cdots.
\end{equation}
Here $\mathcal{F}=\{\Phi,\phi,f,\bar{f}\}$. Take this expressions into the integrations \eqref{barphhi} and \eqref{ffbar1}, we can get  their relationships, which are very useful when we make numerical computations. At the first order, we see that,
\begin{equation}\label{ffbarv0}
\begin{split}
    &\bar{f}(u,v)|_{r=0}=f(u,v)_{r=0}=F_1(u),\\
    &\Phi(u,v)|_{r=0}=\phi(u,v)|_{r=0}.
    \end{split}
\end{equation}
We see that function $F_1(u)$ is free in this system. The fact that the equations of motion cannot  define the function of $F_1(u)$  reflects  the gauge freedom of parameterization of time. By this gauge freedom, we can choose,
\begin{equation}\label{boundary2}
    F_1(u)=1.
\end{equation}
This is done usually in some papers. This choice can lead some convenice when we compute the proper time at the origin point because it is just the coordinate time. However, it is not convenient to find the time and position of black hole formation as we can see later that $f,\bar{f}$ must diverge  when apparent horizon appears. This will break the stability for time-evolution equation of $\Phi$ in Eq.~\eqref{eqscalar2}. In this paper, we will choose the gauge,
\begin{equation}\label{boundary3}
    f(u,v)|_{r=R}=1.
\end{equation}
By the integration for $f$ in \eqref{ffbar1}, it is equivalent to the gauge choice,
\begin{equation}\label{boundary3}
    F_1(u)=\exp\left[-4\pi\int_{v_0}^{v_m(u)}\frac{|\Phi- \phi|^2r'}{r}dv\right].
\end{equation}
Here $v_m(u)$ is defined by the equation $r(u,v_m(u))=R$. When black hole begins to form, $f$ and $\bar{f}$ are both bounded, so that  Eq.~\eqref{eqscalar2} can be stable in numerical calculation. But in this gauge choice, the proper time at the origin point is no longer the coordinate time. In fact, it is given by,
\begin{equation}\label{proptime}
    \tau(u)=\int_0^u\left.\sqrt{f(\tilde{u},v)\bar{f}(\tilde{u},v)}\right|_{r=0}d\tilde{u}=\int_0^uF_1(\tilde{u})d\tilde{u}.
\end{equation}
The black hole will form when $u\rightarrow\infty$, but the proper time at the  origin point $\tau$ is finite.

At the finite boundary $r\rightarrow R$, we need not to specify conditions for metric functions. As the same as in Ref.~\cite{Maliborski:2012gx}, we impose the Dirichlet's condition for $\phi$  as,
\begin{equation}\label{boundaryphi}
    \phi(u,v)|_{r=R}=0.
\end{equation}

In this system, to perform the evolution, we need to give out the initial values of $r(0,v)$ and $\Phi(0,v)$. There is no any requirement for them except that $r(0,v)$ is monotonous with respect to $v$. We will set,
\begin{equation}\label{initx}
r(0,v)=v.
\end{equation}
This setting does not lose the generality, because for any initial setting $r(0,v)=h(v)$, we always can redefine the $v$ such as $v\rightarrow h(v)$ which makes no difference in physics. The initial value $\Phi(0,v)$ will be specified latter when we make numerical calculation.

\subsection{Trapped surface and apparent horizon}
The trapped surface and apparent horizon can be obtained by standard steps. The null tangent vector field for out-going null geodesics congruence is,
\begin{equation}\label{nullvector1}
    k_\mu dx^\mu=-\bar{f}du.
\end{equation}
However, it is not  in its form of affine parameter. It can be found that,
\begin{equation}\label{kdk1}
    k^\mu\nabla_\mu k_\nu=\kappa k_\nu
\end{equation}
with
\begin{equation}\label{kappa1}
    \kappa=\frac{\bar{f}'}{fr'}.
\end{equation}
%
%To find the  null tangent vector field for out-going null geodesics congruence in its affine form, we define a new null vector field as,
In fact for any function $a(u,v)\neq0$, the vector,
\begin{equation}\label{nullvector2}
    \tilde{k}_\mu =a(u,v)k_\mu,
\end{equation}
is also a null tangent vector field for out-going null geodesics congruence. The expansion for out-going null geodesics congruence tangent to $\tilde{k}_\mu$ is,
\begin{equation}\label{expn1}
    \theta=\nabla^\mu \tilde{k}_\mu-a\kappa=\frac{2\bar{f}a}{fr}.
\end{equation}
To remove the coordinate singularity at the origin point  $r=0$, we can choose $a=r/2$. This is just for convenience  when we make numerical computations.

The initial apparent horizon is given by $\theta=0$, i.e., $\bar{f}/f=0$. From the expression in ~\eqref{ffbar1}, the expansion is independent of the gauge choice for $F_1(u)$. So different gauge choices for $F_1(u)$ can give same apparent horizon. If we choose gauge $F_1(u)=1$, then $f(u,v)>1$ and $\bar{f}(u,v)>1$. So when apparent horizon forms, $f(u,v)$ and $\bar{f}$ must be infinite. This will lead that the equation for $\Phi$ in Eq.~\eqref{eqscalar2} is unstable. As  a ressult  this gauge choice is bad for finding the black hole formation. For the gauge choice in Eq. \eqref{boundary3}, we see that condition $\bar{f}/f=0$ means that $f(u,v)\rightarrow0$ and $\bar{f}(u,v)\rightarrow0$. Then the equation for $\Phi$ in Eq.~\eqref{eqscalar2} is stable, which might be a better  choice for performing numerical computation. However, in this gauge choice, we can never really reach the case $\bar{f}/f=0$, because the evolution of system will become slower and slower measured by null time $u$. In practice, we can set a threshold value for $\theta$.

\section{Numerical setup}
\label{Numsim}
\subsection{Initial data}
\label{Numsetup}
In this section we discuss our numerical solver in details. Especially, we will introduce how to deal with the timelike boundaries in the double null coordinates. We will also  make some numerical checks to show the precision of our numerical solver.

The numerical results that we will present arise from a special family of initial value as,
\begin{equation}\label{initialv}
    \Phi(0,v)=\epsilon v^4\exp\left[-\sigma_1^{-2}\tan^2(\frac{v\pi}{2R})+iv/(\sigma_2R)\right].
\end{equation}
with four parameters $\epsilon,\sigma_1,\sigma_2$ and charge $e$. Because there is a symmetry with $e\rightarrow-e$ in the system, we set $e\geq0$. Our solver is based on MATLAB R2010b. The main code and some simple guidance or examples can be downloaded from Ref. \cite{Mcode}.

\subsection{Timelike boundaries}
A numerical simulation of gravitation collapse in an asymptotically flat case in such null coordinates was first performed by Goldwirth and Piran in Ref. \cite{Goldwirth} and improved by Refs.~\cite{Gundlach,Garfinkle} to study the scaling behavior of  the black hole mass. Here we will mainly follow  the algorithm proposed in Ref.~\cite{Garfinkle}.  But, in previous studies,  one needs not to add any boundary condition except for at  the origin point, which is not the case discussed in this paper. We improve the algorithm in Ref.~\cite{Garfinkle} so that it can be suitable for our perfectly reflecting boundary conditions at $r=R$. Our algorithm is as follows.

For the initial value $r(0,v)=v$ and $\Phi(0,v)$, we first discretize the spatial coordinate $v$ into $\{v_0,v_1, \cdots, v_N\}$ by equal interval $\Delta v$ with $v_0=0$ and $v_N=v_m=R$ and obtain the values of $r(0,v)$ and $\Phi(0,v)$ at these grid points $\{r_0(0), r_1(0), \cdots, r_N(0)\}$ and $\{\Phi_0(0), \Phi_1(0), \cdots, \Phi_N(0)\}$. Except for the first 4 grid points of $\phi, f$ and $\bar{f}$,  we use composite 4th order finite difference to compute  $r'$ and Simpson's rule to compute integrations in \eqref{barphhi}, \eqref{charger}, \eqref{guageA} and \eqref{ffbar1}, respectively.
The first few grid points involve  small $r$, which may cause big error and cannot be numerically integrated directly. Instead, we compute them by the 3rd order Lagrange interpolation. For example, for any integration,
\begin{equation}\label{exampleint}
    H=\int_{v_0}^vhr'dv=\int_0^{r(v)}hdr,
\end{equation}
we express $h$ at first 4 grids as,
\begin{equation}\label{interp1}
    h=h_0+h_1r+h_2r^2+h_3r^3
\end{equation}
by the 3rd order Lagrange interpolation using the values of $r$ and $h$ at these grids. Then the value of integration \eqref{exampleint} at these grids are obtained by this formula,
\begin{equation}\label{interp2}
    H=h_0r+h_1r^2/2+h_2r^3/3+h_3r^4/4.
\end{equation}
Further we obtain the time-evolution equations for $r$ and $\Phi$ at the spatial grid at this moment\footnote{Here we say `this moment' because the boundaries of $r=0$ and $r=R$ are evolutive with respect to null time $u$ and spatial grid points are different at different time.}. By 4th order explicit fixed step Runge-Kutta method, we can obtain the next step of $\{x_0(\Delta u), x_1(\Delta u), \cdots, x_N(\Delta u)\}$ and $\{\Phi_0(\Delta u), \Phi_1(\Delta u), \cdots, \Phi_N(\Delta u)\}$. To keep the numerical stability, the 6th order dispersion term is implied. Theoretically, our algorithm has 4th convergence in spatial difference.

Before going to the next step, we have to be careful to deal with the boundary conditions. As once a null trajectory arrives at the origin $r=0$, it bounces and disperses along $u$=const to the boundary of $r=R$. As in the null coordinates, `time' is denoted by $u$, this bounce  will occur at an instantaneous time. Once the wave packet disappears at  $r=0$, a new wave packet will appear at the boundary of $r=R$. The grid point is therefore lost when the light ray hits the origin. So we first need to check the values in $\{r_0(\Delta u), r_1(\Delta u), \cdots, r_N(\Delta u)\}$ and delete the grid points corresponding to $r<0$. Let $r_i$ is the first point such that $r_i>0$. We directly delete all the grids $v<v_i$. On the other hand, as $r_N(\Delta u)<R$, the boundary of $r=R$ now is not covered by spatial grid points. We need to find position of $r=R$ and may need to add some points into the spatial grid. To do so, we use the last a few grids (the number should be larger than 5) to obtain an approximately  polynomial expression $r=\mathcal{R}(v)$ and $v=\mathcal{V}(r)$ by 3rd order Lagrange interpolation. Then the boundary $r=R$ will locate at,
\begin{equation}\label{vcvalue}
    v=v_m=\mathcal{V}(R),
\end{equation}
If $v_m-v_N<\Delta v$, we don't need to add any point. Otherwise, we need to add some new points such that $\{v_{N+1},v_{N+2},\cdots,v_{N+k}\}$ and corresponding values of $r$ such as $\{r_{N+1}(\Delta u), r_{N+2}(\Delta u), \cdots, r_{N+k}(\Delta u)\}$ into the grid. Here $k$ is the integral part of $(v_m-v_N)/\Delta v$ and $v_{N+j}=v_N+j\Delta v, r_{N+j}(\Delta u)=\mathcal{R}(v_{N+j})$. The values of $\phi$ at these additional points are also obtained by 3rd order Lagrange interpolation at the grid points $\{v_{N-2},v_{N-1},v_N,v_m\}$ and their  values of $\phi$, $\{\phi_{N-2}(\Delta u),\phi_{N-1}(\Delta u),\phi_N(\Delta u),0\}$. Here we have used the boundary condition $\phi(u,v)|_{x=R}=0$. With the relation $\Phi=\phi+r\phi'/r'$, we can get the values of $\Phi$ at these added points.

\begin{figure}
\begin{center}
\includegraphics[width=0.35\textwidth]{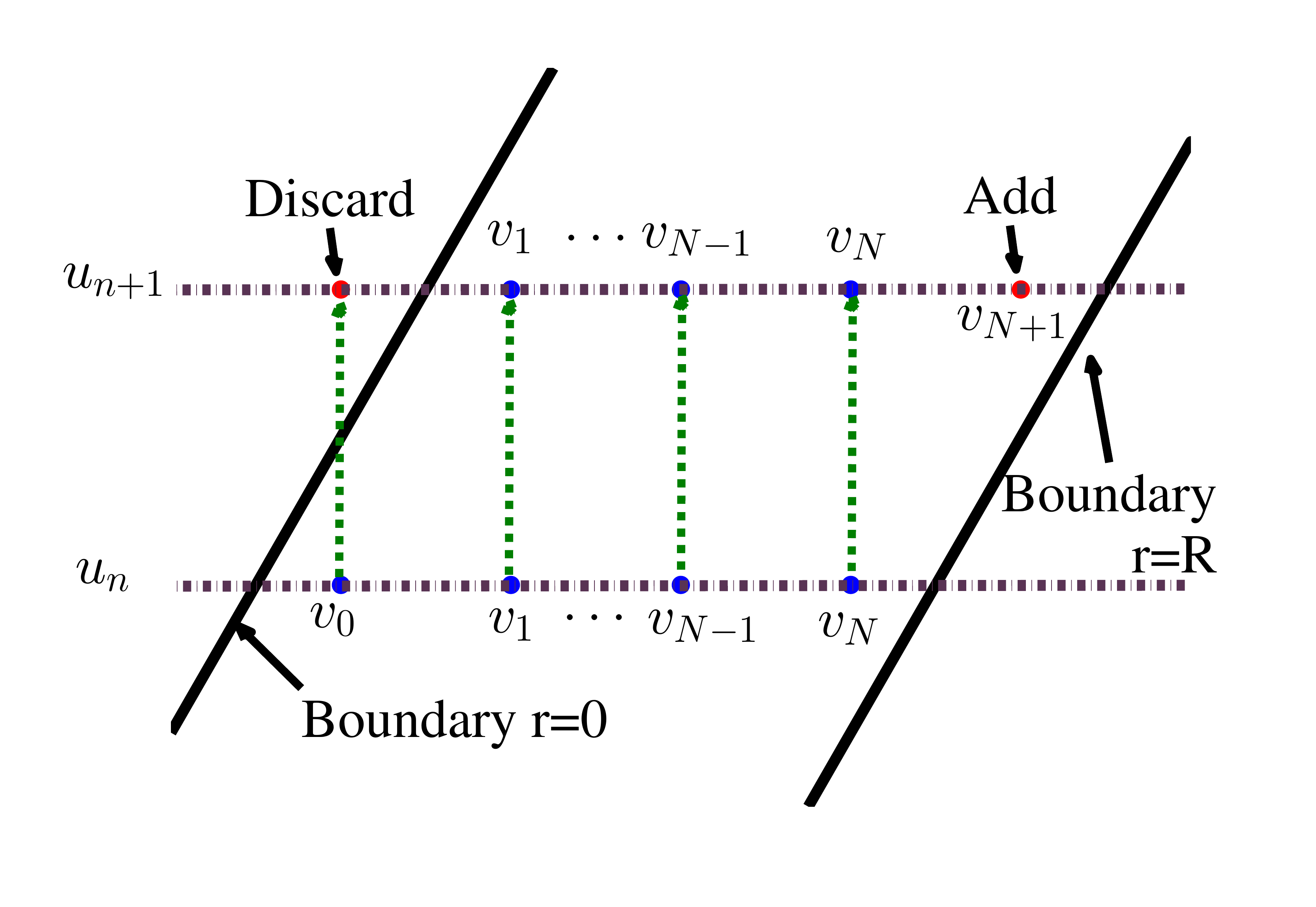}
\caption{The schematic diagram for the process to deal with boundary conditions. $v_m$ is the  value  of $v$ which gives $r=R$. The values of $\phi$ at the new added points can be obtained by 3rd order Lagrange interpolation by using the grid points $\{v_{N-2},v_{N-1},v_N,v_m\}$ and the values of  $\phi$, $\{\phi_{N-2}(\Delta u),\phi_{N-1}(\Delta u),\phi_N(\Delta u),0\}$.}
\label{bound1}
\end{center}
\end{figure}

Now we obtain the grids $\{v_i, v_{i+1}\cdots,v_{N+k}\}$ at the new time and the function values of $r$ and $\Phi$ at these grid points such as $\{r_i, r_{i+1}\cdots, r_{N+k}\}$ and $\{\Phi_i, \Phi_{i+1}\cdots, ,\Phi_{N+k}\}$. Then we use them as the initial values of next step. A schematic diagram for process of discarding and adding points is shown in Fig.\ref{bound1}.
To be accurate, the time step $\Delta u$ is determined so that in each step the number of added points is no more than 1, so we need,
\begin{equation}\label{restrcdu}
    \Delta u\leq2\Delta v.
\end{equation}

\begin{figure}
\begin{center}
\includegraphics[width=0.5\textwidth]{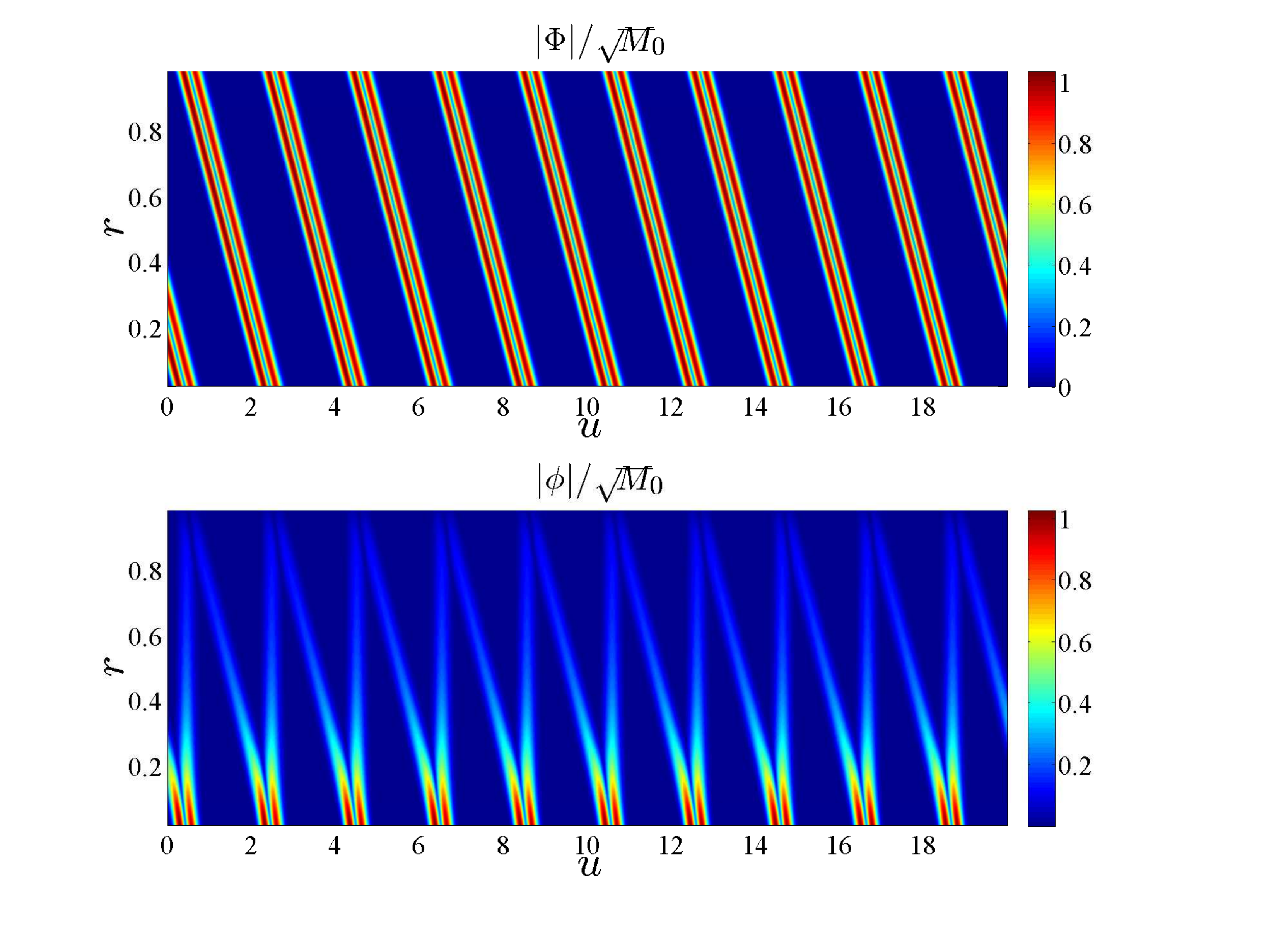}
\caption{A typical example for the value of $|\Phi(u,v)|/\sqrt{M_0}$  and $|\phi(u,v)|/\sqrt{M_0}$ in $r-u$ plane when black hole does not appear. The initial value we take is shown in Eq.~\eqref{initialv} with $\epsilon=30, 2\sigma_1=\sigma_2=1/2, R=e=1$.}
\label{rtp}
\end{center}
\end{figure}

This process is iterated as many times as necessary until the minimum value of $\bar{f}/f$ is less than a threshold value which suggests that an apparent horizon appears and black hole begins to form. An example of numerical solution for $\phi$ normalized with the total mass $M_0$ is shown in Fig.~\ref{rtp}. It is clearly seen how the bounce happens when the wave packet arrives at the origin point and a new wave packet appears at the boundary of $r=R$.

\subsection{Tests}
In this subsection, various tests will be made to show the correctness and accuracy of our numerical solver as well as the consistency of the finite difference method used to evolve the system. We use a variety of diagnostic tools, including monitoring of the additional equation \eqref{chackerror1}, the conserved quantities and convergence tests of the primary dynamical variables.

\begin{figure}
\begin{center}
\includegraphics[width=0.23\textwidth]{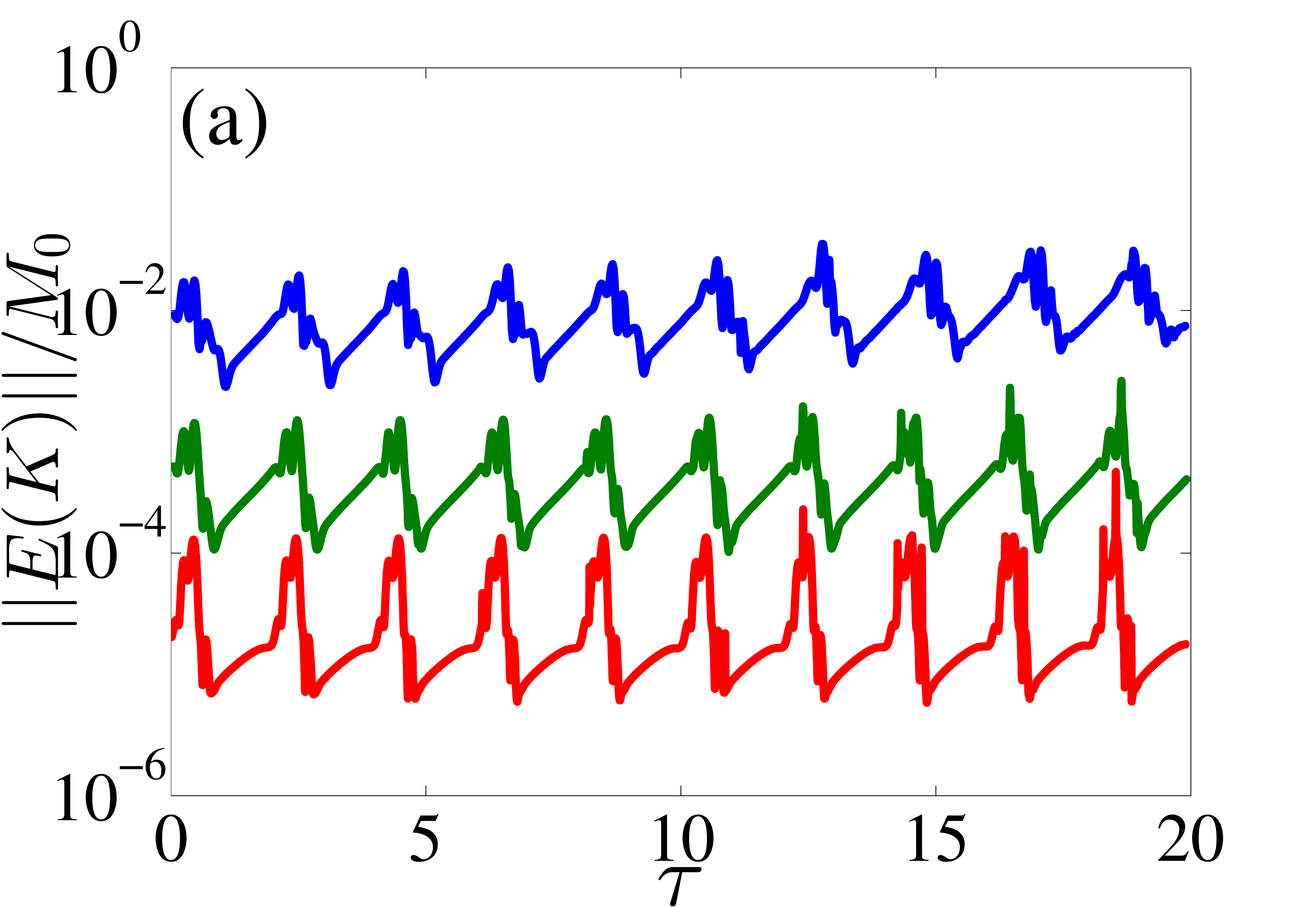}
\includegraphics[width=0.23\textwidth]{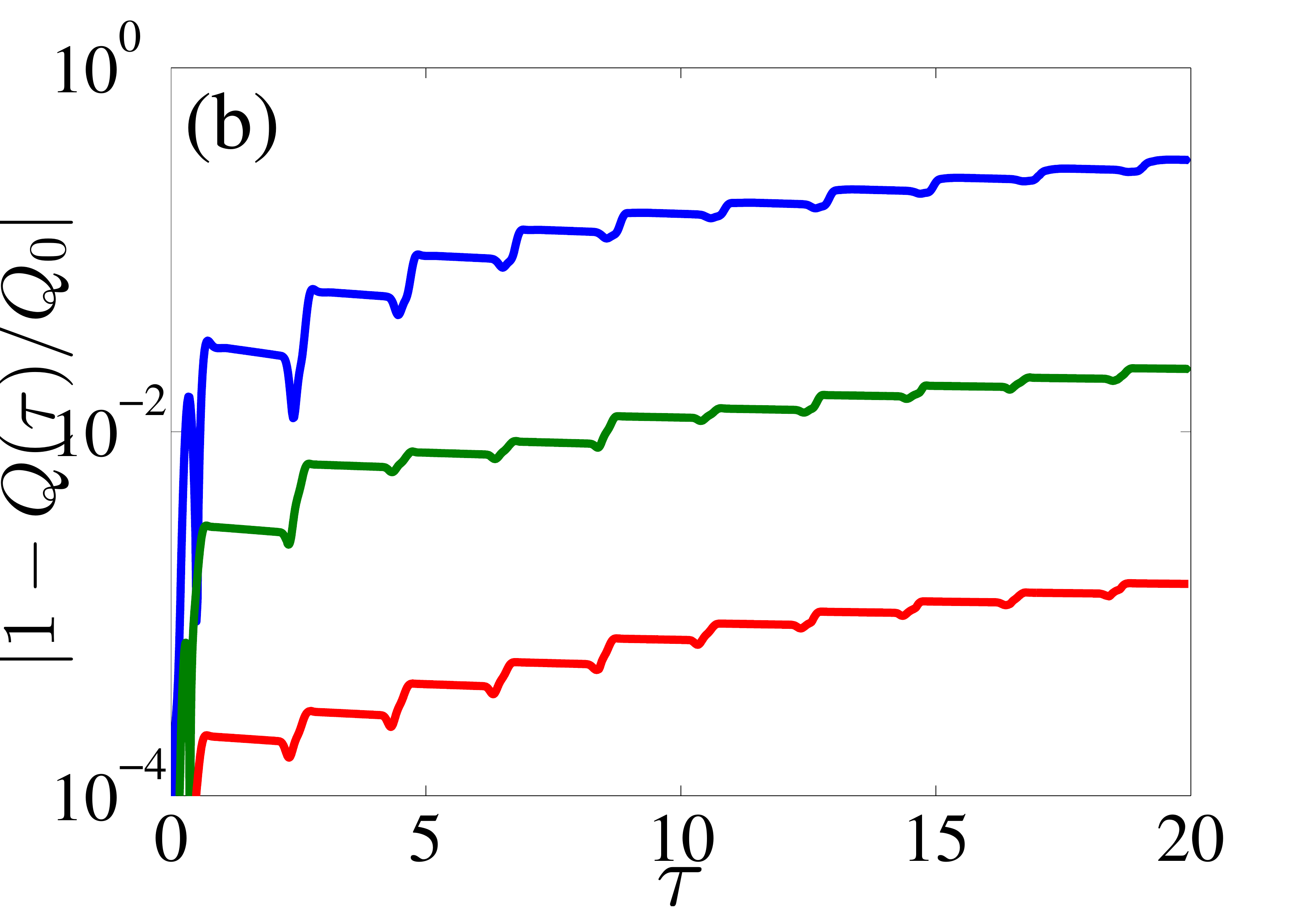}
\includegraphics[width=0.23\textwidth]{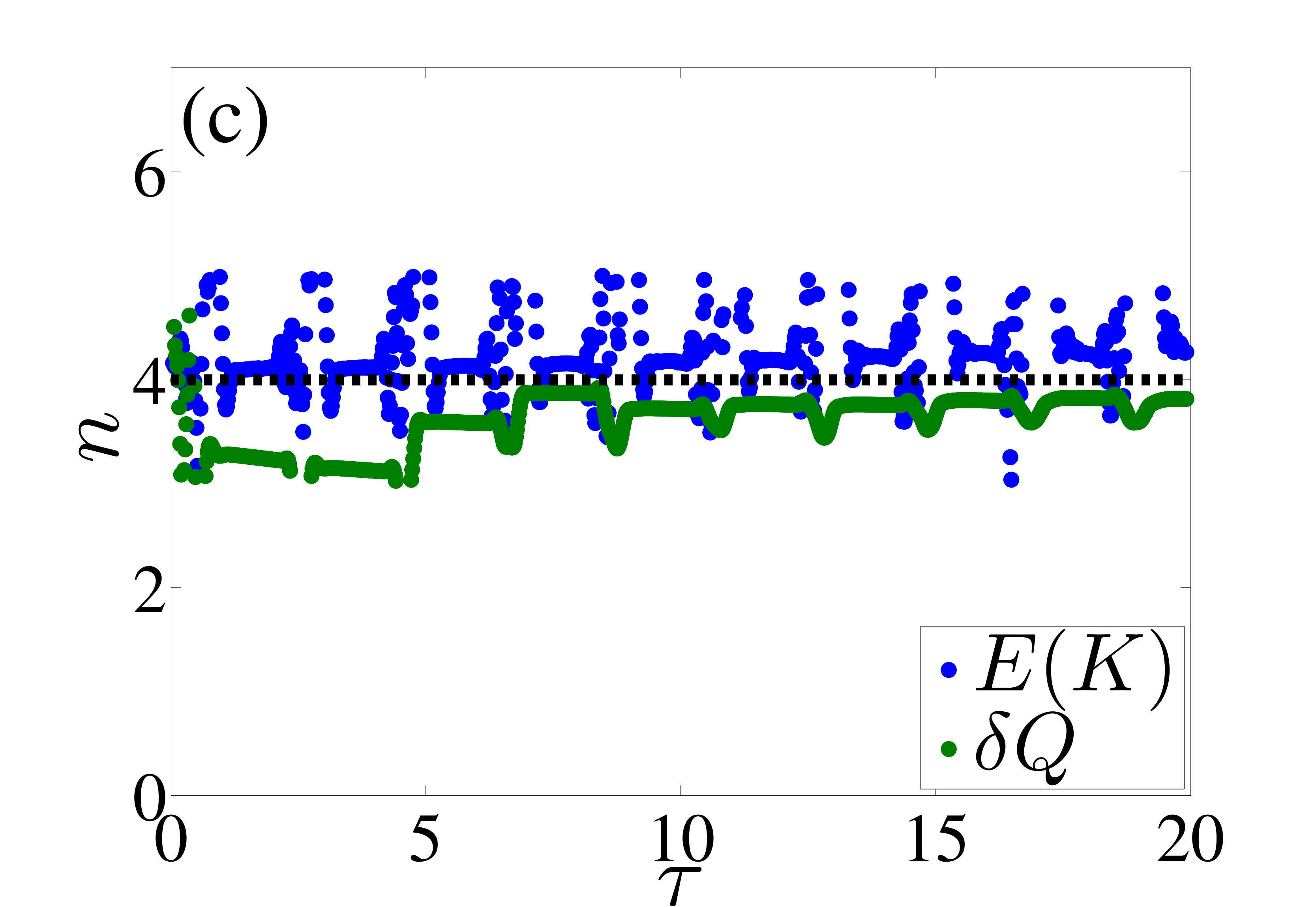}
\includegraphics[width=0.23\textwidth]{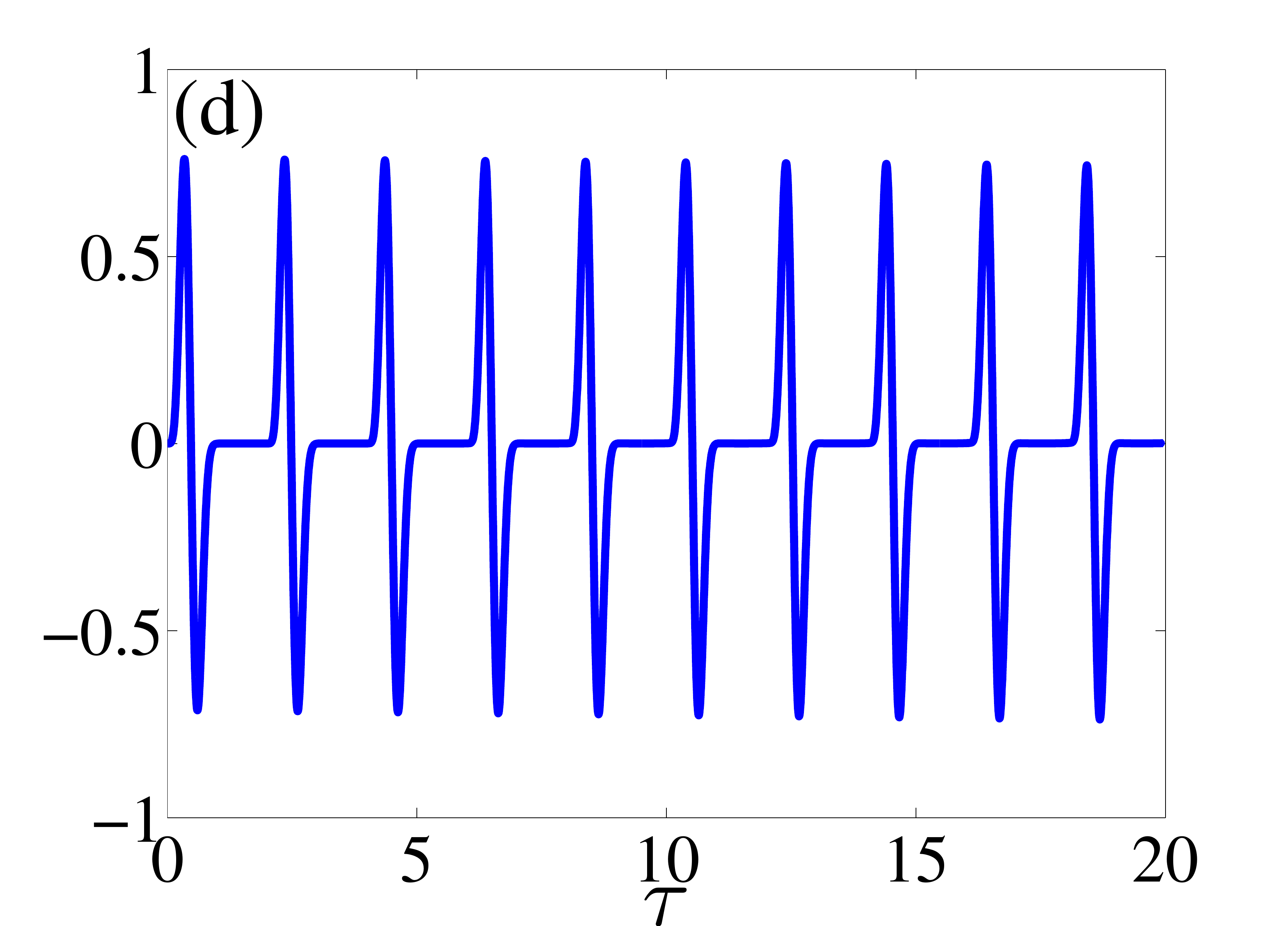}
\caption{(a): The  root mean square value of the error for $uu$-component of Einstein's equations at different proper time at the origin point. We normalize them with the total mass $M_0\simeq3.6\times10^{-4}$. The spatial step length for blue line, green line and red line are 1/100, 1/200 and 1/400 (from upper to down), respectively. (b): The error for total charge with respect to proper time at the origin point. The spatial step lengths for blue line, green line and red line are 1/100, 1/200 and 1/400  (from upper to down), respectively. (c): The convergence test for the $uu$-component of Einstein's equations and the total charge. The spatial step length $h$ is 1/200. (d): The values of Re$\phi$ at the origin point with respect to proper time at the origin point. The initial value we take is shown in Eq.~\eqref{initialv} with $\epsilon=10, 2\sigma_1=\sigma_2=1/2, R=e=1$.}
\label{errK}
\end{center}
\end{figure}

In order to monitor the evolution of  Eq.\eqref{chackerror1}, we define the $l_2$ norm for any function $\mathcal{F}(v)$ as,
\begin{equation}\label{chack1}
    E(\mathcal{F})=\parallel \mathcal{F}\parallel.
\end{equation}
Here $\parallel\cdot\parallel$ means the   root mean square  value at the $v$-grids. A typical result for the error of $uu$-component of Einstein's equations $E(K)$ at different proper time of origin point is shown in Figs.~\ref{errK}(a). This is the  most important indicator for the accuracy in our algorithm.
The total charge $Q_0$ contained inside the mirror is a conserved quantity. The charge contained within the sphere of radius $r$ is obtained directly from integration \eqref{charger} by solving the system and $Q_0=Q(u,v)_{r=R}$. The deviation of the total charge from its initial value is defined as,
\begin{equation}\label{errcharge}
    \delta Q(\tau)=|1-Q(\tau)/Q_0|.
\end{equation}
The typical values of $\delta Q(\tau)$ for different spatial step length are plotted in Figs.~\ref{errK}(b). This is another indicator for the accuracy in our algorithm.

The convergence test should also be considered. It is more convenient to choose $E(K)(u)$ and $\delta Q(\tau)$ as the test functions since the exact values of them are both zero.  Let $\mathcal{F}(u)$ stand for $E(K)(u)$ or $\delta Q(\tau)$, then a discrete approximation to it computed at finite difference resolution $\Delta v=h$ is denoted by $\mathcal{F}(u;h)$. Fixing initial data, if the computation is $n$-th order convergent, we can expect that when $h$ is small enough,
\begin{equation}\label{definedC}
    \mathcal{F}(u;h)=h^n\mathcal{F}_1(u)+\mathcal{O}(h^{n+1}).
\end{equation}
We perform a sequence of calculations with resolutions $h$ and $h/2$ and then compute a convergent order as,
\begin{equation}\label{convergC}
    n=\log_2[\mathcal{F}(u;h)/\mathcal{F}(u;h/2)].
\end{equation}
From  Figs.~\ref{errK}(c), we see that $n=3\sim5$, which provides clear evidence that the solver is better than the third order convergence throughout the time evolution.

Theoretically, our numerical algorithm should have 4th order convergence. The deviation arises in handling the boundary conditions. This can be seen by comparing  Figs.~\ref{errK}(a-c) with (d), which shows that the peaks appearing in  Figs.~\ref{errK}(a-c) always happen at the time when the wave packet arrives at the origin point. This can be understood as follows. To obtain the function values at the boundary $r=R$, we use the 3rd Lagrangian interpolation, which gives the 3rd global accuracy. When the wave packet reaches the origin point or just appears at the boundary of $r=R$, because of 3rd Lagrangian interpolation, the accuracy is just 3rd order \footnote{Higher order interpolations will lead instability because of the  Runge's phenomenon.}. When the wave packet is away from the boundary of $r=R$, the effects caused by the handling of boundary conditions are negligible as the functions of $\phi$ and $\Phi$ are  near zero in the region near $r=0$ or $r=R$. Then the error is dominated by the algorithms of spatial integration and time evolution, which gives nearly a 4th order convergence. Generally speaking, our algorithm shows a 3rd convergence.

As we introduced in  section~\ref{Introd}, one of main advantages using the double null coordinates is that it has higher efficiency when a black hole forms. The reason why the double null coordinates are more efficient  than the usual Schwarzschild coordinates can be understood as follows. In the usual Schwarzschild spherical coordinates $\{t,r,\theta,\phi\}$ with the metric form taken in Ref.~\cite{Maliborski:2012gx}, the fields will change along $r$-direction rapidly  in a very small spatial region near the horizon. To obtain a precise  result, we have to make the step be so small that there are enough grids in this narrow region. However, in our null coordinates, though the spatial step in the null-spatial direction $v$ is still uniform, the physical distance is highly non uniform. In fact, as the apparent horizon begins to form, the grids will be more and more concentrated  on the place where apparent horizon begins to form so that the  changes of fields  with respect to $v$ are still smooth.
\begin{figure}
\begin{center}
\includegraphics[width=0.42\textwidth]{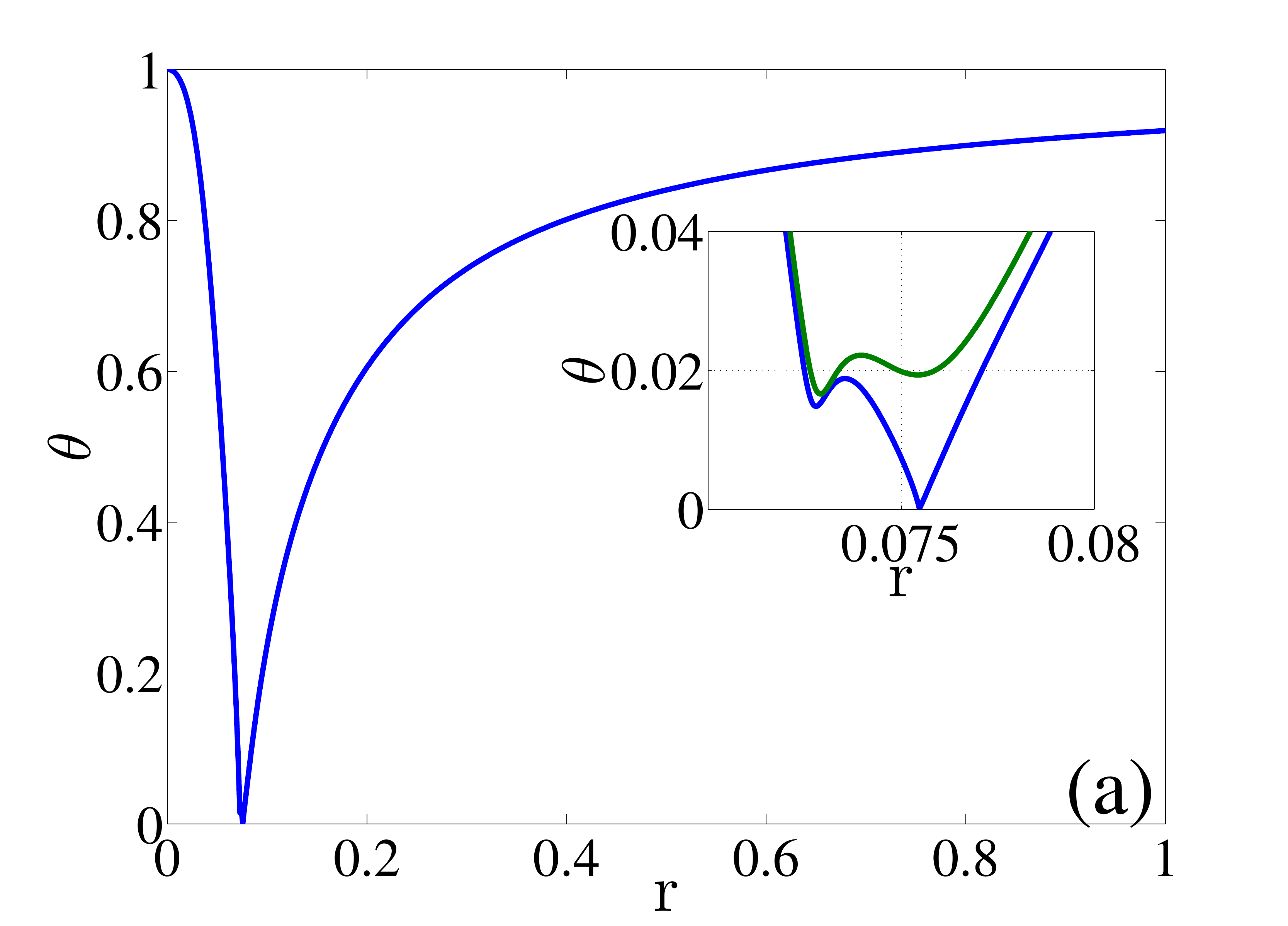}
\includegraphics[width=0.42\textwidth]{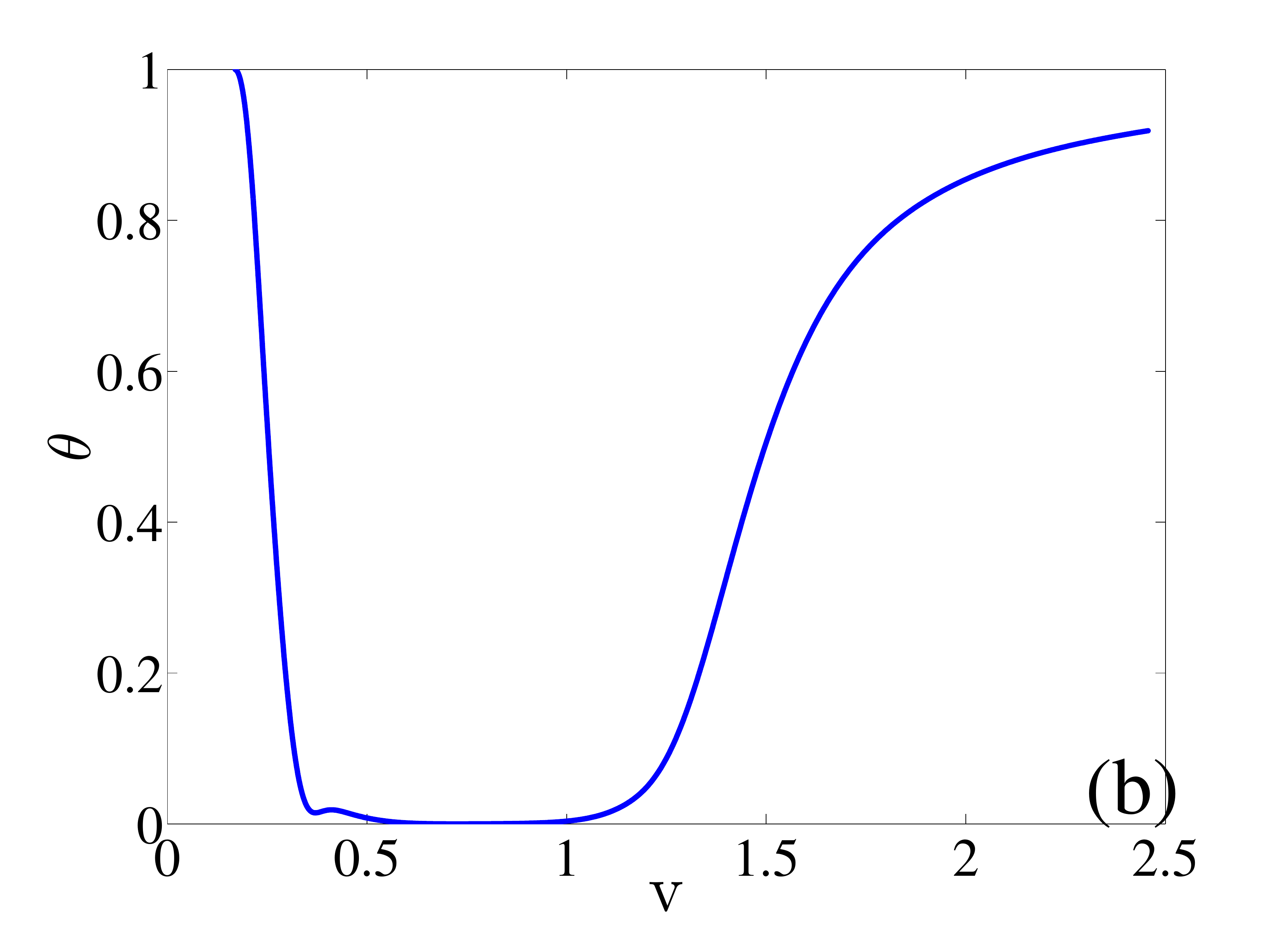}
\caption{The expansion $\theta$ in the $r$-grid and null spatial $v$-grid. In inset  of (a), the green line is the case with the threshold value $\theta=0.015$, while the blue line is the case with the threshold value $\theta=10^{-4}$. The initial value we take is shown in Eq.~\eqref{initialv} with $\sigma_1=\sigma_2=1/4, R=1, e=0$ and $\epsilon=340$.}
\label{gtheta}
\end{center}
\end{figure}

 Fig.~\ref{gtheta} is a typical example showing a comparison between  the expansions of $r=$constant null surface in  the usual Schwarzschild spherical coordinates and  double null coordinates. From Fig. \ref{gtheta}(a), we see that we have to set the  threshold value of expansion is small enough so that we can get a reasonable value of $r_h$ ( comparing the green line and blue line in the inset of Fig. \ref{gtheta}(a)). However, when $\theta\rightarrow0$, the expansion will have a very narrow peak. To cover such sharply changing region, we have to make the $r$-grid refined enough so that there are enough grids in this narrow region. In uniform grids, one needs at least $10^5$ order in the number of  grid points in spatial direction. On the other hand, in the double  null coordinates, we see that there is no any peak and the expansion near the horizon is smooth. To obtain the same accuracy, we need only about $10^3$ grids in the spatial direction.

\section{Numerical simulations}
\label{results}
\subsection{Multiple critical collapse}
\begin{figure}
\begin{center}
\includegraphics[width=0.38\textwidth]{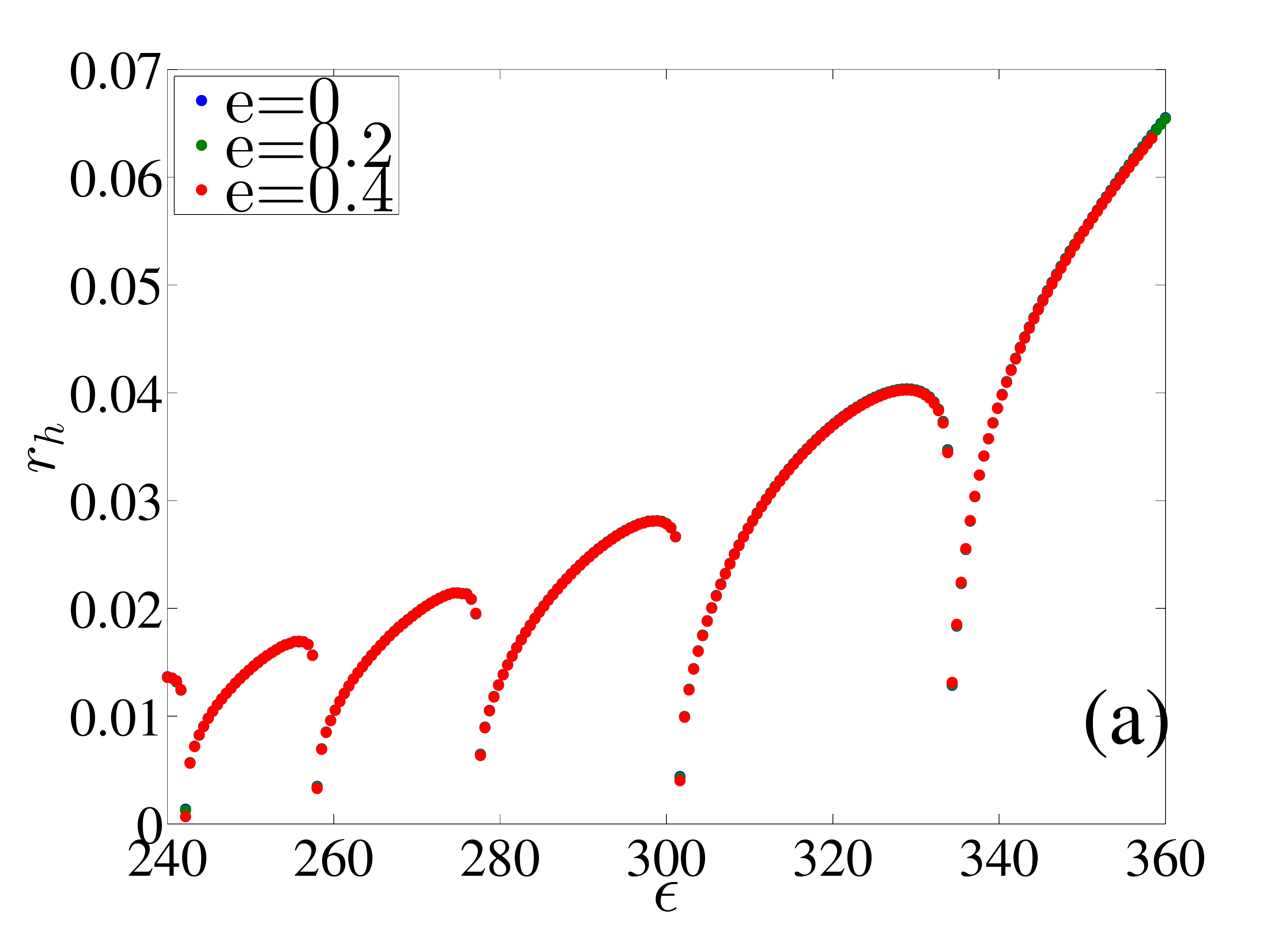}
\includegraphics[width=0.38\textwidth]{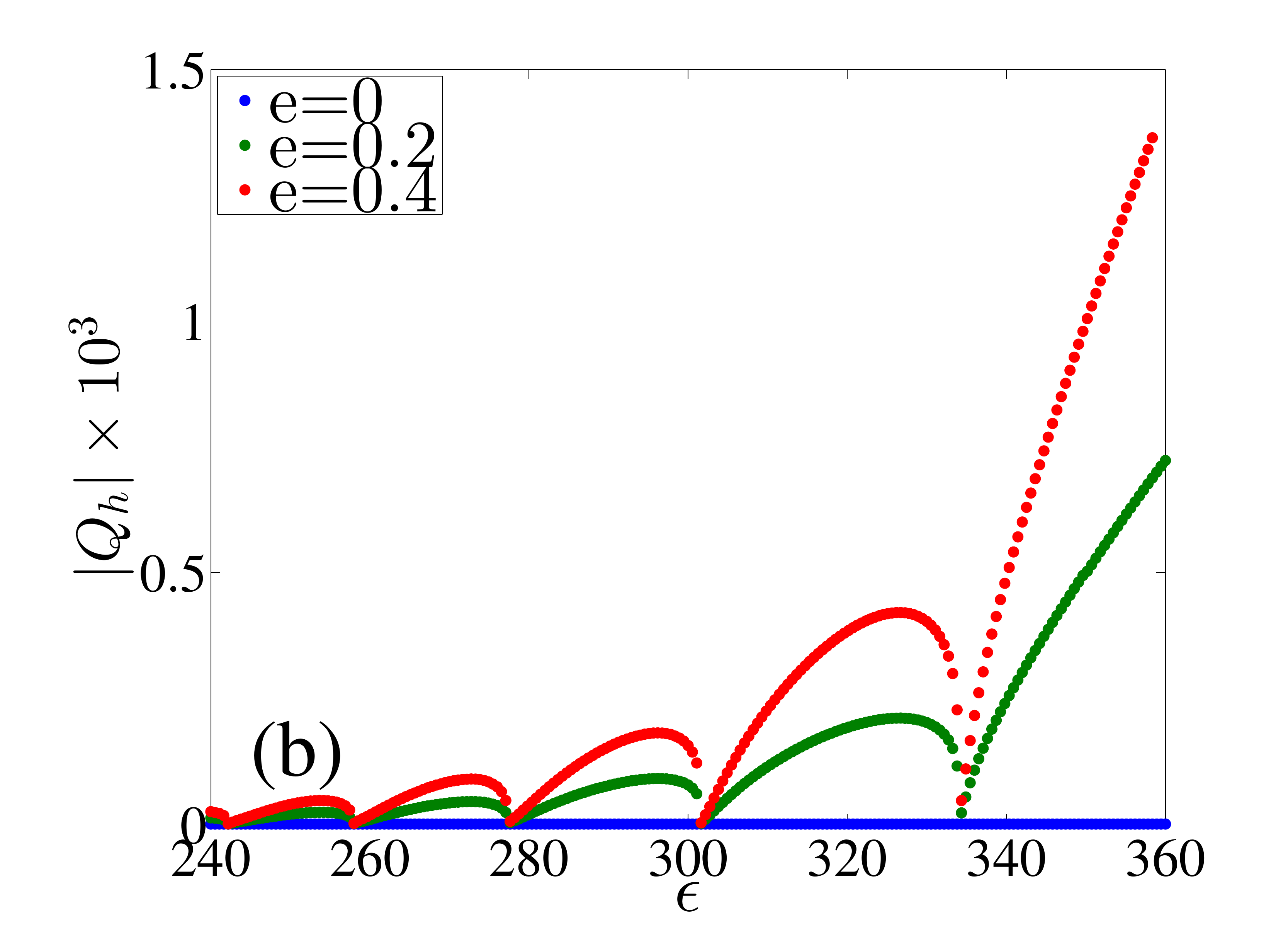}
\includegraphics[width=0.38\textwidth]{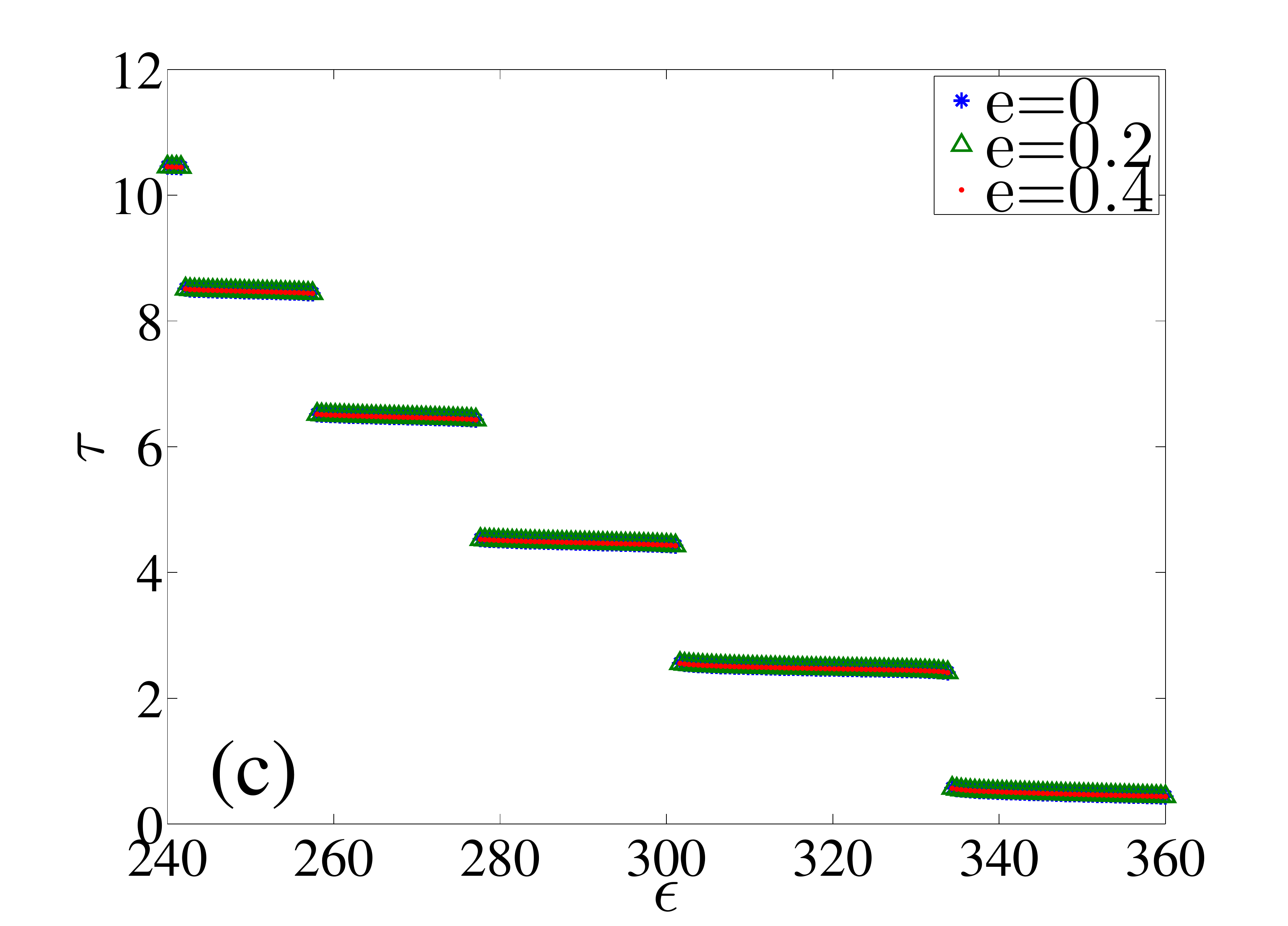}
\includegraphics[width=0.38\textwidth]{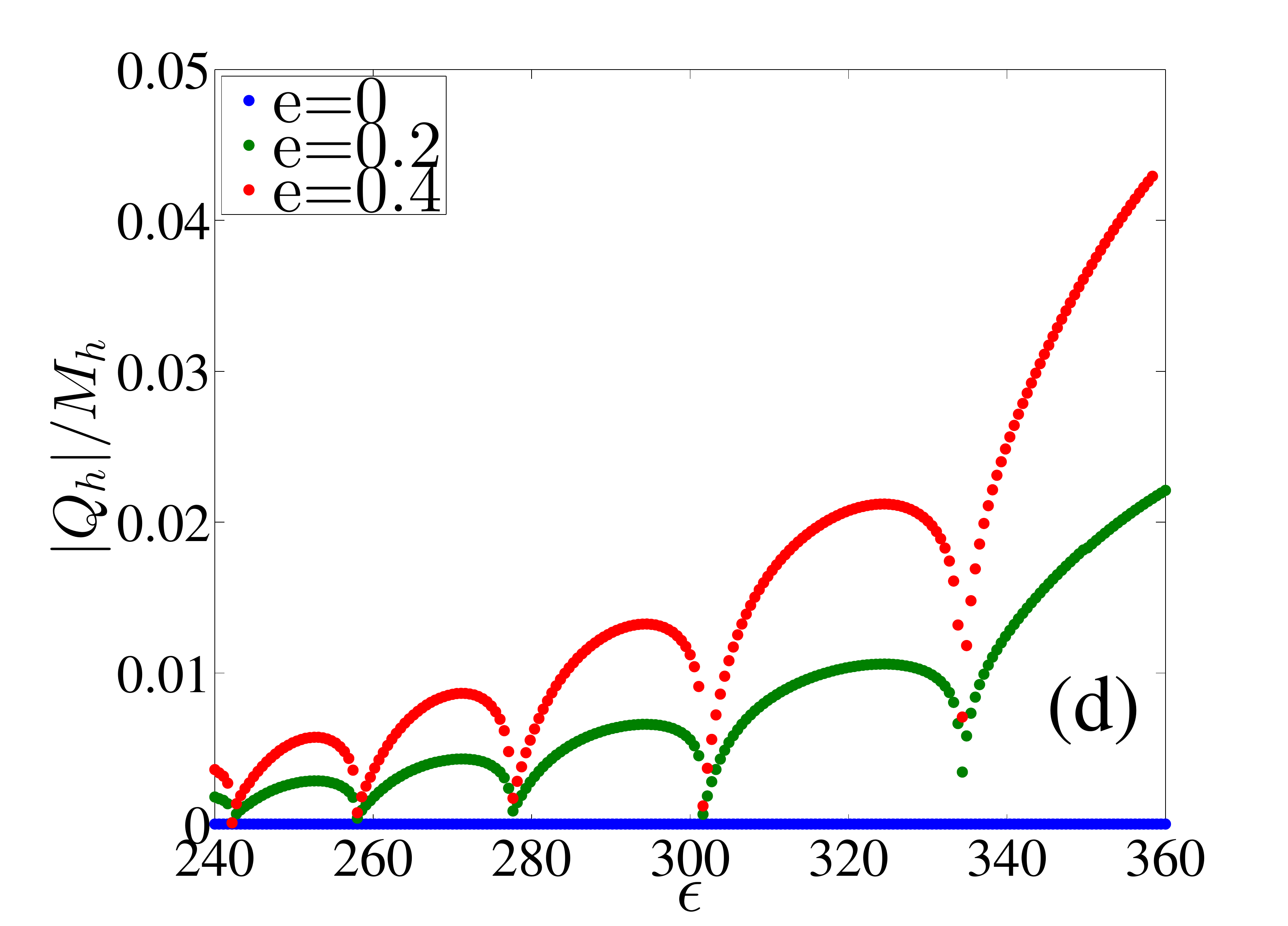}
\caption{(a): The mass $M_h$ contained within the initial apparent horizon  vs amplitude $\epsilon$ with  different charge $e$. (b): The charge contained in initial apparent horizon. Note that there are gaps when amplitudes tend to critical values from the subcritical direction. (c): The proper time $\tau$ at the origin point when  the black hole forms  with respect to  amplitude $\epsilon$ in different charge $e$. The lines of $\tau$ for different $e$ are nearly coincident with other. (d): the ratio  between the mass and charge contained in the initial apparent horizon. Similar to the case in (c), there are gaps when amplitudes tend to critical values from the subcritical direction. The initial value we take is shown in Eq.~\eqref{initialv} with $\sigma_1=\sigma_2=1/4, R=1$.}
\label{colla1}
\end{center}
\end{figure}

From this subsection, we will present our numerical results. A complex scalar with an additional self-interacting potential  in a fixed AdS background with Maxwell field has been studied in Ref.~\cite{Liebling:2012gv}.  Here in our setup, the Maxwell field and gravitational fields are both dynamical. We will first study the black hole formation in this system, i.e., the strong field case. Here we fix $\sigma_1=\sigma_2=1/4, R=1$. Our main results are shown in Fig.~\ref{colla1}.

In the case of $e=0$, our model is just the one with two massless scalar fields minimal coupling with gravitational field, so the results are very similar to what have been obtained in Ref.~\cite{Maliborski:2012gx}. We see that there are a series of critical amplitudes $\epsilon_n$($n=0,1,2$,$\cdots$) which give zero initial mass for black hole and the horizon formation time exhibits jumps of size $\Delta\tau\simeq2$. This coincidence  is also the  evidence to show our solver and algorithm are effective and accurate. However, our solver and algorithm have an obvious advantage. In our double null coordinates, we just take about 3 hours to obtain all the data in Fig.~\ref{colla1}. This advantage can release us from long time waiting for the numerical results  and gives us more time to try different cases, which can make us obtain more guidances from numerical results to understand physics behind.

Our interest of course  is  the case with $e\neq0$. In  Fig.~\ref{colla1}(a) and (c), we plot the values of initial black hole mass $M_h$ and the formation time $\tau$ at the  origin point with respect to $\epsilon$ when $e=0, 0.2$ and $0.4$, respetively. In our scanning region for $\epsilon$ and the initial family \eqref{initialv}, the black hole will always form and the influence of charge is just to increase the initial black hole mass. Our numerical results show that the critical amplitudes are dependent of the value of $e$ very weak and  the lines for different charge are nearly coincident with each other. In very step of Fig.~\ref{colla1}(c), the formation times for different charges are also very close. For larger $e$, because of the strong repulsive force coming from Maxwell field,  the change of expansion $\theta$ near the horizon formation varies very slowly, we can't confirm if a black hole forms.

Except for the initial horizon radius, the initial charge $Q_h$ contained in the apparent horizon when a black hole forms is also a very important quantity. From  Fig.~\ref{colla1}(b), we see that it has a similar behavior as the initial black hole mass. As the amplitude approaches to its critical value from the supercritical direction,  $Q_h$ also tends to zero. As the values of charge and mass contained in the initial apparent horizon  both tend to zero near the critical amplitudes, the ratio of them is also an interesting quantity. In  Fig.~\ref{colla1}(d), we plot the value $Q_h/M_h$ for different charges and amplitudes. There is a jump behavior with respect to amplitude. Our numerical results show that $Q_h/M_h$ tends to zero as the amplitude tends to the critical value from the supercritical direction. This means that the black hole charge tends to zero faster than its mass and the initial black hole will more and more neutral. This result has also been found in the same system without the reflecting mirror in Ref.~\cite{Hod:1996ar}. Our results imply that the reflection boundary condition does not change this behavior.

\begin{figure}
\begin{center}
\includegraphics[width=0.43\textwidth]{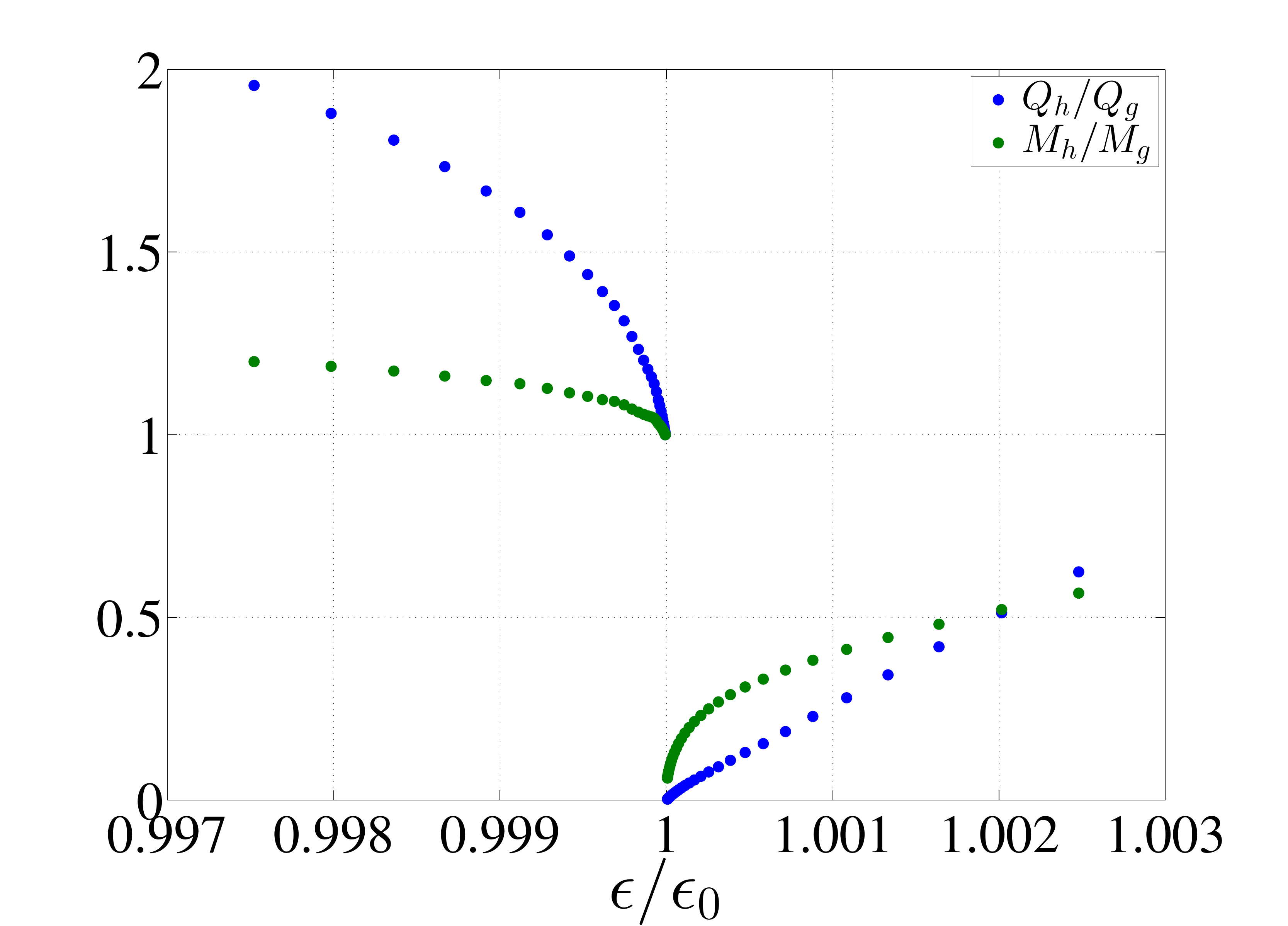}
\caption{The mass and charge contained in the initial apparent horizon when $\epsilon$ is  near the first critical value $\epsilon_0$. Here $Q_g\approx1.18\times10^{-4}$ and $M_g\approx1.56\times10^{-2}$. The initial value we take is shown in Eq.~\eqref{initialv} with $\sigma_1=\sigma_2=1/4, R=1, e=0.3$.}
\label{gap1}
\end{center}
\end{figure}

In the recent work in Ref.~\cite{Olivan:2015fmy}, a mass gap for critical gravitational collapse in asymptotically AdS space-time was found by numerical simulation. We find that this gap can also appear in the asymptotically flat space-time with perfectly reflecting boundary. In addition, there is also a gap for charge contained in the initial apparent horizon. Our results imply that such a gapped behavior for the critical solutions is a universal phenomenon in the closed system in gravitational collapse.  In Fig.~\ref{gap1}, we plot typical results of the mass and charge contained in the initial apparent horizon for the first critical solution.
%However, our algorithm failed to give the scaling exponents at the  lefthand of critical solutions.
%

\subsection{Critical exponents at different critical points}
As we have shown, the double-null coordinates are very suitable for studying the behavior near the critical amplitudes. In this subsection, we will pay attention to the scaling behaviors near the critical amplitudes.

The first group of scaling relations is on the mass $M_h$, and charge $Q_h$ with respect to the tuning parameter $\epsilon$ in initial family and charge $e$, which can be characterized by following three relationships as,
\begin{equation}\label{scaling1}
\left\{
\begin{split}
    &M_h|_{e=0}\propto(\epsilon-\epsilon_{n0})^{\beta_n^+}, ~~\epsilon\rightarrow\epsilon_{n0}^+,\\
    &\lim_{e\rightarrow0}Q_h/e\propto(\epsilon-\epsilon_{n0})^{\gamma_n^+},~~\epsilon\rightarrow\epsilon_{n0}^+,\\
    &M_h|_{\epsilon=\epsilon_{n0}^+}\propto|e|^{2-\alpha_n^+}, ~~e\rightarrow0, ~\text{and if}~\epsilon_n<\epsilon_{n0}\\
    \end{split}
    \right.
\end{equation}
Here index $n=0,1,2,3,\cdots$, which stands for  the different critical amplitudes. $\epsilon_{n}$ is the n-th critical amplitude for fixed charge $e$ and $\epsilon_{n0}$ is the n-th critical amplitude when $e=0$. The third scaling relationship in \eqref{scaling1} exists only when $\epsilon_n<\epsilon_{n0}$.

\begin{figure}
\begin{center}
\includegraphics[width=0.42\textwidth]{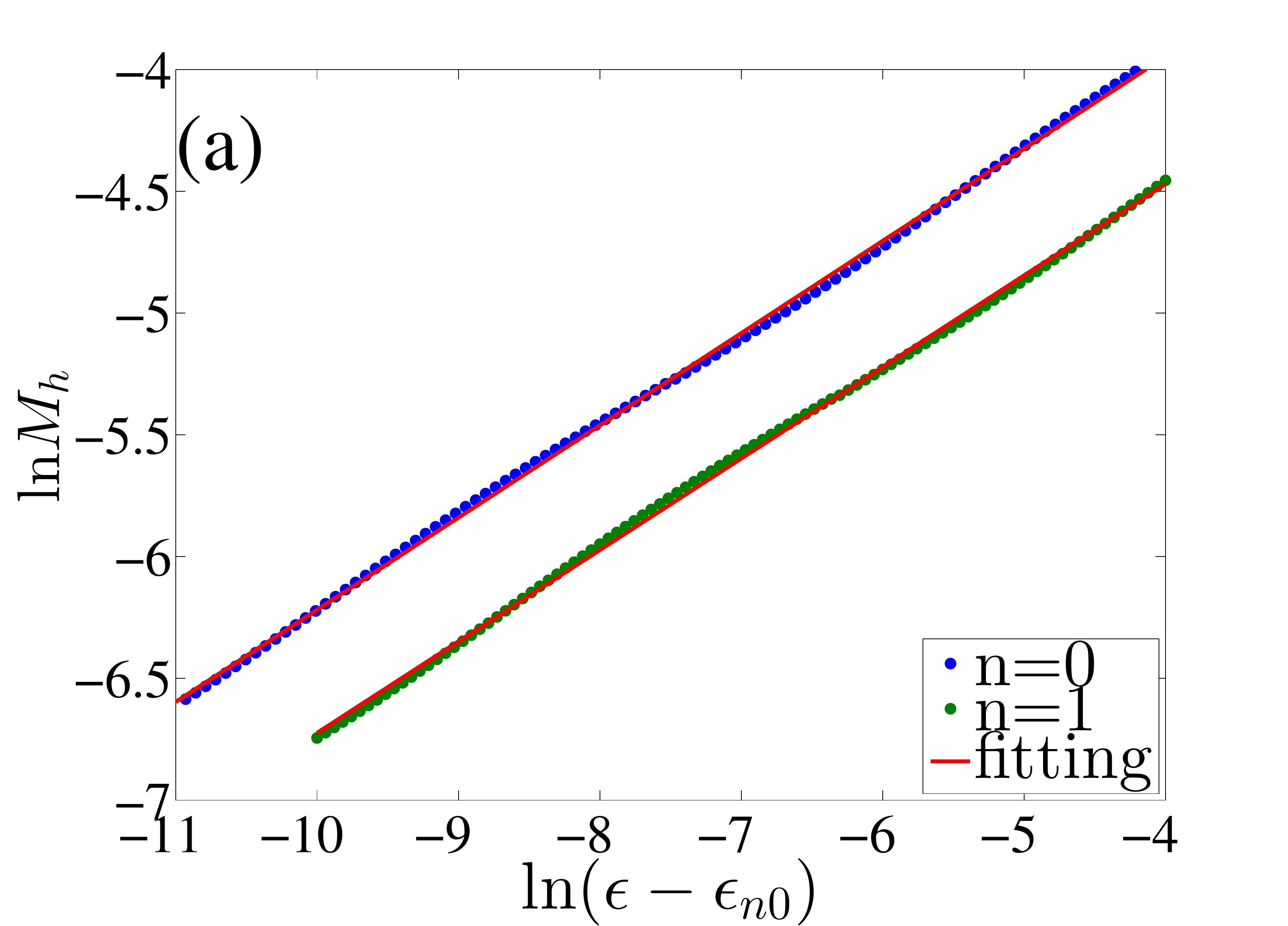}
\includegraphics[width=0.42\textwidth]{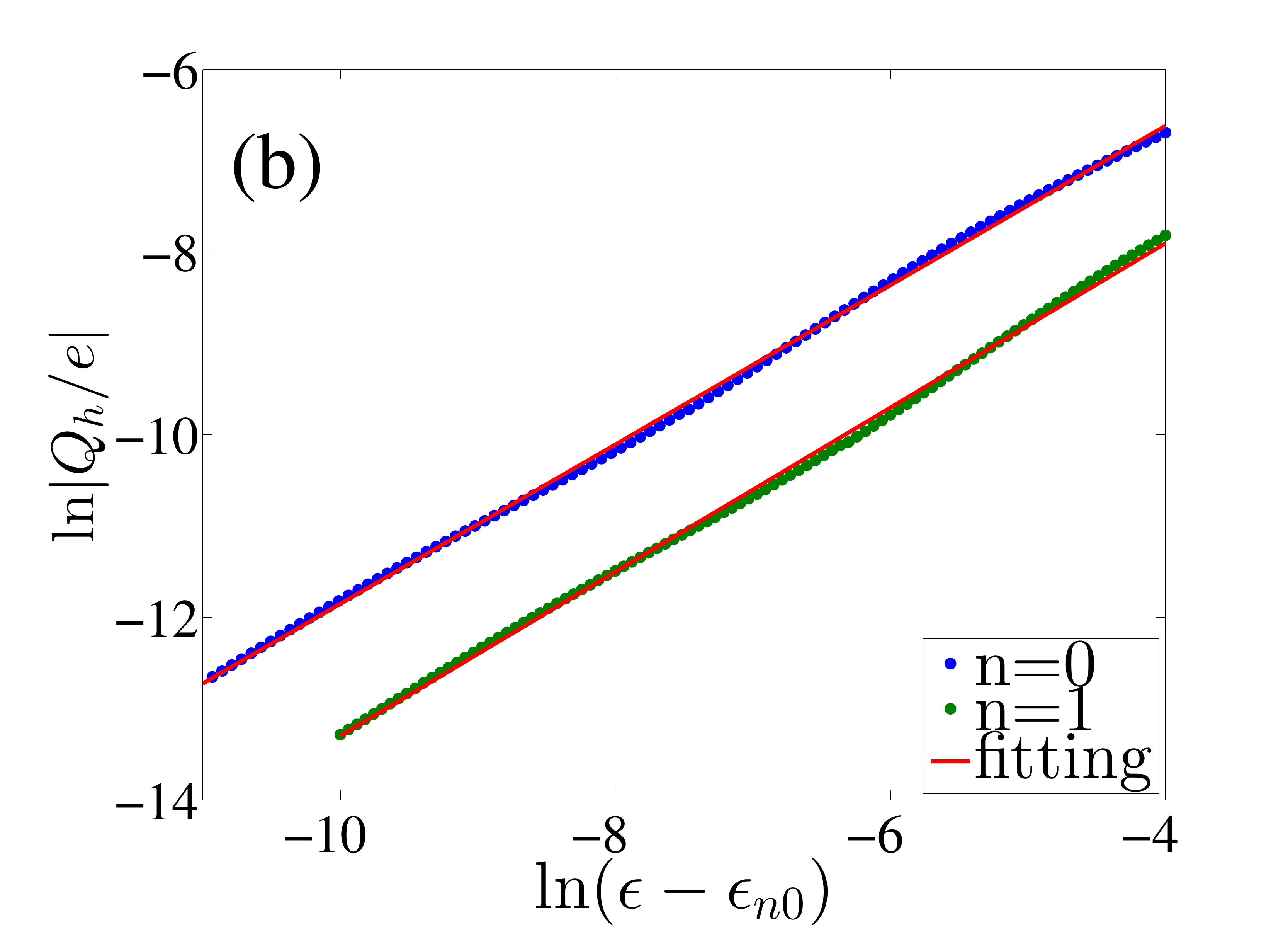}
\includegraphics[width=0.42\textwidth]{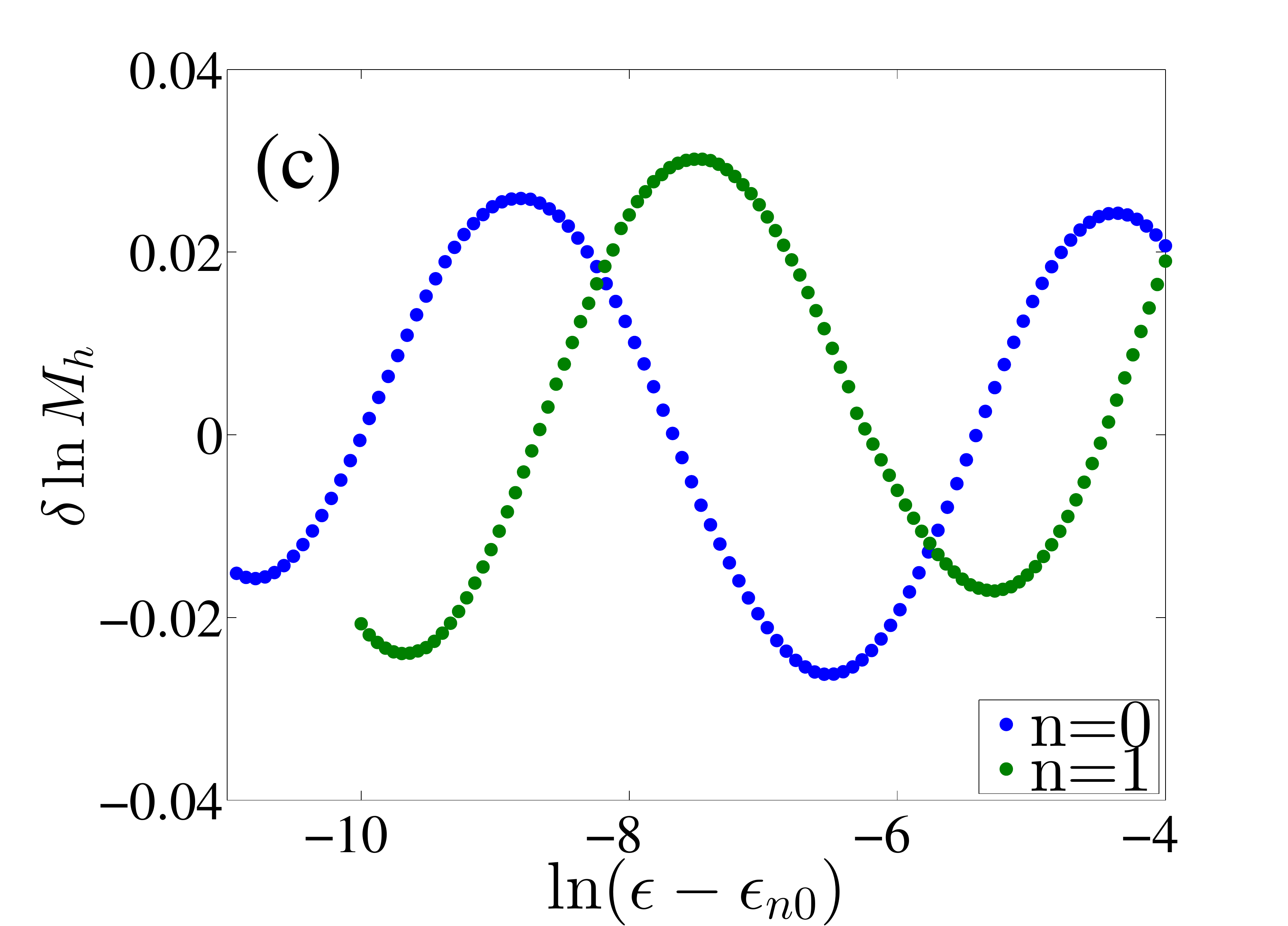}
\includegraphics[width=0.42\textwidth]{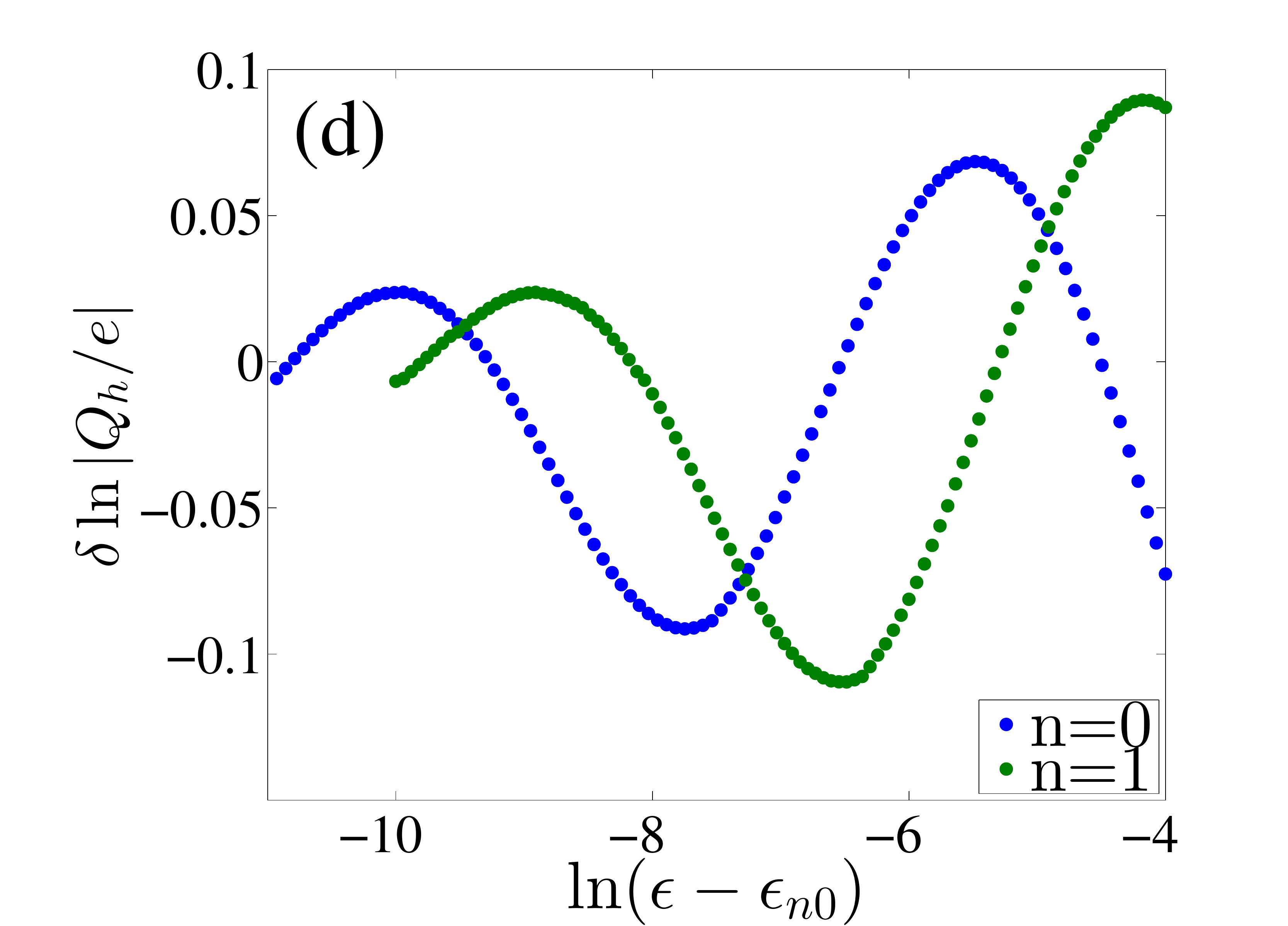}
\caption{(a): The mass contained in the initial apparent horizon vs amplitude $\epsilon$ when $\epsilon\rightarrow\epsilon_{n0}^+$ for $n=0$ and $1$. (b): The charge contained in the initial horizon vs amplitude $\epsilon$ when  $\epsilon\rightarrow\epsilon_{n0}^+$ for $n=0$ and $1$. (c): The periodic structure of the deviation in mass scaling relation. (d): The periodic structure of the deviation in charge scaling relation.}
\label{colla2}
\end{center}
\end{figure}
\begin{figure}
\begin{center}
\includegraphics[width=0.42\textwidth]{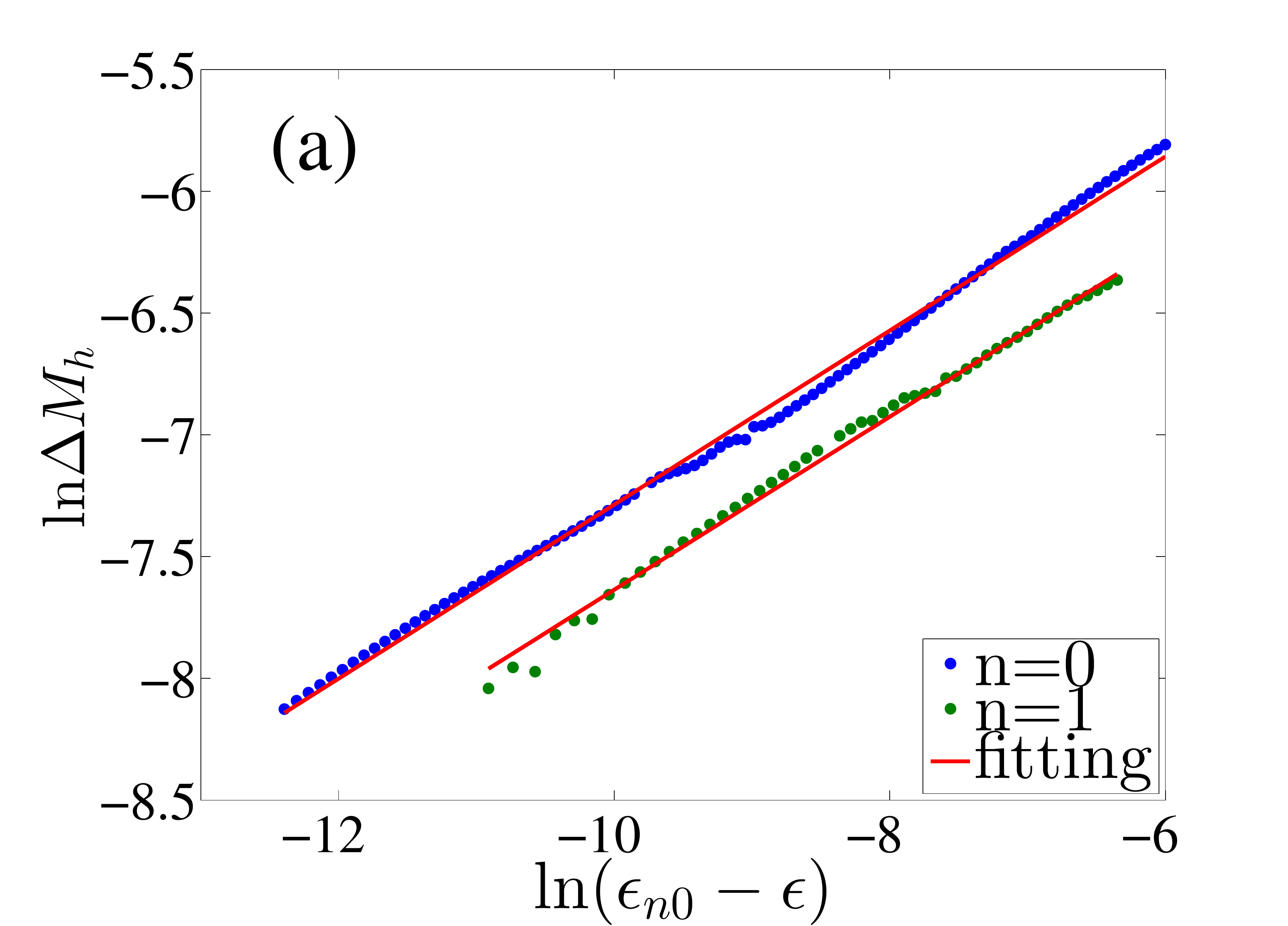}
\includegraphics[width=0.42\textwidth]{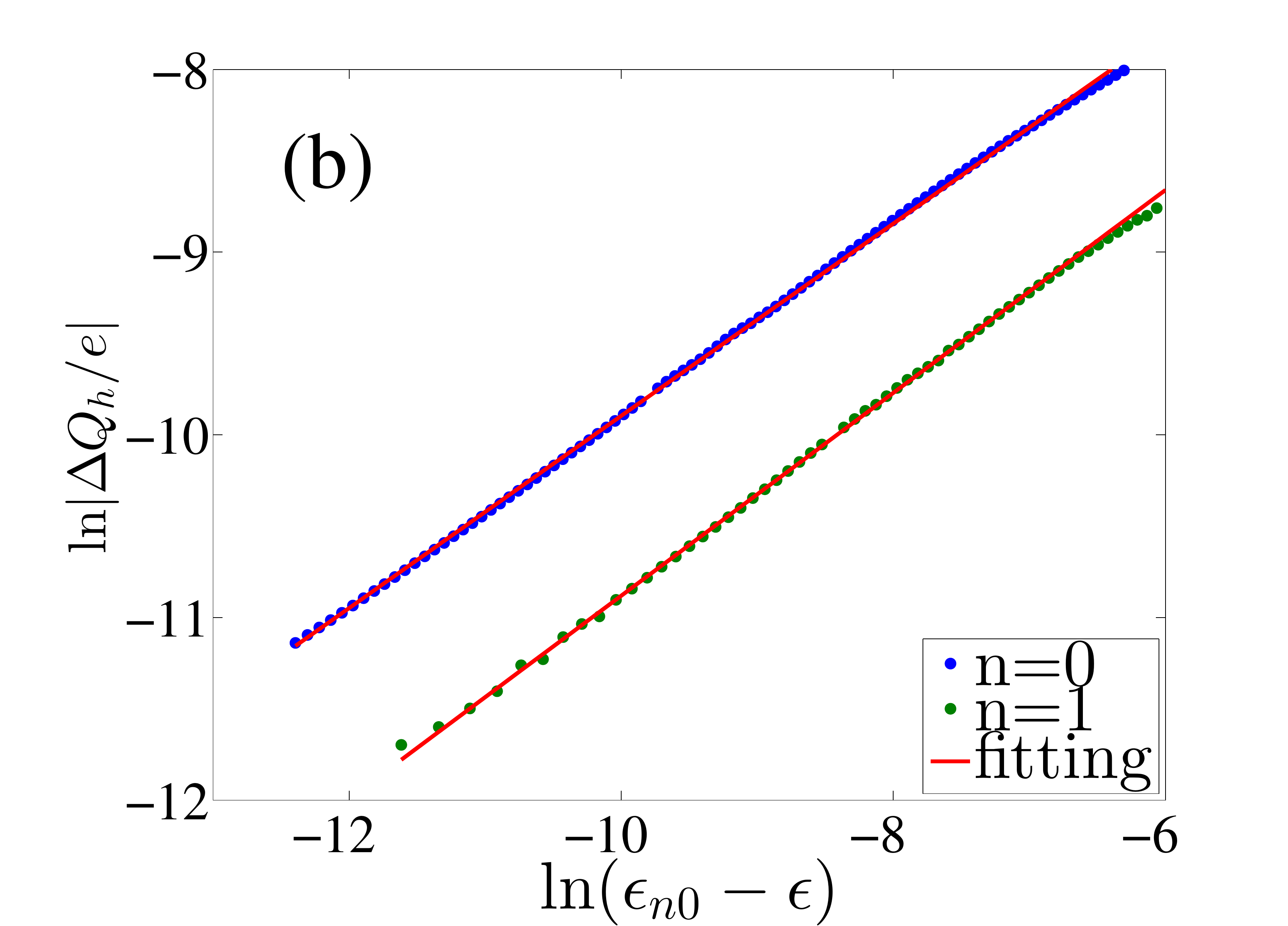}
\caption{(a): The mass contained in the initial apparent horizon vs amplitude $\epsilon$ when $\epsilon\rightarrow\epsilon_{n0}^-$ for $n=0$ and $1$. (b): The charge contained in the  initial horizon vs amplitude $\epsilon$ when  $\epsilon\rightarrow\epsilon_{n0}^-$ for $n=0$ and $1$. Here $\Delta M_h=M_h-M_{ng}$ and $\Delta Q_h=Q_h-Q_{ng}$}
\label{colla3}
\end{center}
\end{figure}

In the case of asymptotically flat spacetime without reflection mirror, there is only one critical solution. When the reflection mirror is imposed,  there are multiple critical solutions so that the critical exponents may be different at different critical amplitudes. But after $n$ reflections from the mirror boundary, locally asymptotic Choptuik's solution \cite{Choptuik} will form because critical solution only depends on the local properties near the origin point. Then from the discussions in Ref.\cite{Choptuik} and Refs.\cite{Gundlach:1996vv,Hod:1996ar}, we find that indeed the critical exponents $\beta_n^+\approx0.37$ and $\gamma_n^+\approx0.88$, independent of $n$. Except for the four scaling relationships in Eqs.~\eqref{scaling1}, other two scaling relations were first proposed in Ref. \cite{Cai:2015jbs} to characterize the influence of interaction on critical amplitudes and formation time of black hole. In the current case, they are,
\begin{equation}\label{scaling2}
\begin{split}
    &\chi_{\epsilon_n}=\partial\epsilon_n/\partial e\propto|e|^{\delta_n},~~e\rightarrow0\\
    &\chi_{\tau_n}^+=\lim_{e\rightarrow0}\frac{\partial\tau}{\partial e^2}\propto(\epsilon-\epsilon_{n0})^{-\eta_n^+}, ~~\epsilon\rightarrow\epsilon_{n0}^+.
    \end{split}
\end{equation}

The scaling equations \eqref{scaling1} and \eqref{scaling2} with five critical exponents $\{\beta_n^+, \delta_n^+, \alpha_n^+, \delta_n, \eta_n^+\}$ give out the critical behaviors when $\epsilon\rightarrow\epsilon_{n0}^+$ and $e\rightarrow0$. However, the  five critical exponents are not independent with each other. Based on the result in Ref.~\cite{Cai:2015b}, $\beta_n^+,  \alpha_n^+$ and $\delta_n$ satisfy the following universal relationship,
\begin{equation}\label{scalingrel}
    \beta_n^+(\delta_n+1)+\alpha_n^+=2.
\end{equation}
For a given charge $e$, if we treat the expansion $\theta$ at the origin point as a function of $\epsilon$, then finding critical solution becomes an eigenvalue problem, i.e., finding suitable $\epsilon_n$ such that $\theta|_{r=0}=0$. As the equations of motion for matter field and metric are all smoothly dependent on charge $e$, we can assume that the eigenvalue $\epsilon_n$ is a function of $e$. Taking the symmetry $e\rightarrow-e$ into account, we can make a Taylor's expansion for $\epsilon_n$ as,
\begin{equation}\label{epsilonn}
    \epsilon_n=\epsilon_n^{(0)}+e^2\epsilon_n^{(1)}+\cdots.
\end{equation}
Here $\epsilon_n^{(k)}, k=0,1,2,\cdots$ are independent of $e$. Take \eqref{epsilonn} into the first one of \eqref{scaling2}, we can obtain that,
\begin{equation}\label{deltavalue}
    \delta_n=1.
\end{equation}
Put \eqref{deltavalue} into \eqref{scalingrel}, we can find that $2-\alpha_n^+=2\beta_n^+$. The value of $\alpha_0^+$ was first computed in Ref. \cite{Hod:1996ar} by numerical fitting and showed $2-\alpha_0^+\approx2\beta_0^+$. Here we used scaling \eqref{scalingrel} with \eqref{deltavalue} to  confirm this analytically.  This is  in agreement with our numerical fitting shown in  Table~\ref{Tab1}. However, when $n\geq1$, numerical simulation shows that $\chi_{\epsilon_n}>0$, which means that for any given $e\neq0$, the critical amplitude $\epsilon_n$ is always larger than $\epsilon_{n0}$. So the critical exponent $\alpha_n^+$ does not  exist for $n\geq1$.

In an open system, there is a sub-leading correction appearing in the mass scaling relationship in  Eqs. \eqref{scaling1}. In the asymptoticlly flat space-time with massless real scalar field, such a sub-leading correction is a periodic function of the critical separation $\ln(\epsilon-\epsilon^*)$ with the period $\Omega=\Delta/(2\beta)$ \cite{Hod:1996az}. Here $\Delta$ is the echoing period which shows a discrete self-similar behavior \cite{Choptuik}. Such a periodic structure in the  sub-leading correction for scaling relationships is the performance of discrete self-similarity.  In the closed system, such as the asymptotically  AdS space-time, such a periodic correction for the scaling relationships when $\epsilon\rightarrow\epsilon_{n0}^+$ also appears~\cite{Olivan:2015fmy}.

In  Fig.~\ref{colla2}, we examine the  sub-leading corrections in the mass and charge scaling relationships for the first two critical solutions  when $\epsilon\rightarrow\epsilon_{n0}^+$. In Fig. \ref{colla2}(a) and (b), we plot the fitting results for $\ln M_h$ and $\ln|Q_h/e|$ with respect to $\ln(\epsilon-\epsilon_{n0})$ for $n=0$ and $1$, respectively, which shows that the numerical results obey the scaling relationships \eqref{scaling1} very well. Then in Fig. \ref{colla2}(c) and (d), we plot the deviations from the fitting curves. A  periodic structure can be clearly seen in Fig. \ref{colla2}(c) and (d) both for mass and charge when $n=0$ or $1$. According to the four curves in   Fig. \ref{colla2}(c) and (d), we find that the values of $\Omega$ are very close to each other and $\Delta\approx3.43$, which is  the same as the case without the reflection mirror found in Refs \cite{Choptuik,Garfinkle}.

In the case with $\epsilon\rightarrow\epsilon_n^-$, we can also define a group of critical exponents $\{\beta^-_n, \alpha_n^-, \delta_n^-, \eta_n^-\}$ in a way as,
\begin{equation}\label{scaling3}
\left\{
\begin{split}
    &(M_h-M_{ng})|_{e=0}\propto(\epsilon-\epsilon_{n0})^{\beta_n^-}, ~~\epsilon\rightarrow\epsilon_{n0}^-,\\
    &\lim_{e\rightarrow0}e^{-1}(Q_h-Q_{ng})\propto(\epsilon-\epsilon_{n0})^{\gamma_n^-},~~\epsilon\rightarrow\epsilon_{n0}^-,\\
    &(M_h-M_{ng})|_{\epsilon=\epsilon_{n0}^-}\propto|e|^{2-\alpha_n^-}, ~\text{if}~e\rightarrow0, ~\text{and}~\epsilon_n>\epsilon_{n0}\\
    &\chi_{\tau_n}^-=\lim_{e\rightarrow0}\frac{\partial\tau}{\partial e^2}\propto(\epsilon-\epsilon_{n0})^{-\eta_n^-}, ~~\epsilon\rightarrow\epsilon_{n0}^-.
    \end{split}
    \right.
\end{equation}
Here $M_{ng}$ and $Q_{ng}$ are the gaps of mass and charge contained in the initial apparent horizon for the n-th critical solution when $e\rightarrow0$. The third scaling relationship in \eqref{scaling3} exists only when $\epsilon_n>\epsilon_{n0}$. According to the results in Ref. \cite{Cai:2015b}, there is also a scaling law for $\beta^-_n$ and $\delta_n^-$,
\begin{equation}\label{scalingrel2}
    \beta_n^-(\delta_n+1)+\alpha_n^-=2.
\end{equation}
This means that only one is independent in  $\beta^-_n$ and $\delta_n^-$. As only when $n\geq1$ can $\epsilon_n$ be larger than $\epsilon_{n0}$,  $\alpha_0^-$ does not exist.

To compare with the case with $\epsilon\rightarrow\epsilon_{n0}^+$, we also  examine the deviations in the mass and charge scaling relations for the first two critical solutions, which are shown in  Fig. \ref{colla3}. We find that the critical exponents are different from the case in Fig. \ref{colla2}. In addition, the critical exponent $\beta_n^-\approx0.36$, which is also different from its counterpart in the asymptotically AdS spacetime reported in Ref. \cite{Olivan:2015fmy}, where numerical result shows that $\beta_n^-\approx0.70$. In the gapless case when $\epsilon\rightarrow\epsilon_{n0}^+$, we know that the values of $\beta_n^+$ in the asymptotically  AdS space-time share the same value as in the asymptotically  flat space-time, because the scaling behavior happens in the infinitesimal region ($r\rightarrow0$) and the influence of cosmological constant can be ignored. However, in the gapped case when $\epsilon\rightarrow\epsilon_{n0}^-$, the  influence of cosmological constant cannot be ignored so that the critical behavior and exponent are different.

Different from the case in  Fig. \ref{colla2}, the numerical error is too large to give out the information about the sub-leading term. This is due to the fact that when the amplitude approaches to $\epsilon_n^-$, the singularity will appear in the origin point  (when $\epsilon=\epsilon_n$,  the horizon radius is zero and the singularity appears at $r=0$.), which leads the Taylor's expansion in Eq. \eqref{interp1} loses its accuration.  As the   periodic structure in the sub-leading term is a manifestation of discrete self-similarity, it is worth to improving the  algorithm  so that  the sub-leading terms can be determined. We hope we can report this in the future.

The values of $\eta_n^\pm$ can only be obtained by fitting the formation time of black hole at the critical solutions for different charge $e$. Our results show that $\eta_n^+\approx0.57$ and $\eta_n^-\approx0.91$ , which are also independent of $n$.

In Table~\ref{Tab1}, we show our numerical fitting results by our algorithm  about the five exponents for the first two critical solutions with a given initial family. The results show the agreement with the scaling laws \eqref{scalingrel}.
\begin{table*}
  \centering
  \begin{tabular}{c|cccccccccccc}
    \hline
    \hline
    $n$& $\epsilon_{n}$&$\beta_n^+$&$2-\alpha_n^+$&$\delta_n$&$\gamma_n^+$&$\eta_n^+$&$|Q_{ng}|/e$&$ M_{ng}$&$\beta_n^-$&$2-\alpha_n^-$&$\gamma_n^-$&$\eta_n^-$\\
    \hline
    0&334.003&0.379&0.751&$1.00$&0.872&0.573&$3.78\times10^{-4}$&$1.53\times10^{-2}$&0.357&---&0.526&0.90\\
    1&301.540&0.376&---&$0.958$&0.899&0.575&$1.67\times10^{-4}$&$1.15\times10^{-2}$&0.355&0.713&0.555&0.92\\
    \hline
  \end{tabular}
  \caption{The best fitting values about critical exponents and gaps for first two critical solutions.  The initial value we take is shown in Eq.~\eqref{initialv} with $\sigma_1=\sigma_2=1/4, R=1$. One can see that the results are agreement with the relationships \eqref{scalingrel}.}\label{Tab1}
\end{table*}

\section{Summary}
\label{summ}
In this paper, we proposed a new method based on the double-null coordinates to investigate the multiple critical gravitational collapse  of charged scalar field in closed system and studied the influence of charge on the multiple critical phenomena in gravitational collapse. The background we consider is the asymptotically flat space-time with a perfect reflecting  mirror at a fixed radius  so that the energy cannot disperse into infinity.

Firstly, we gave out the equations of motion of the system, the conditions for apparent horizon and shown how to deal with the time-like boundary conditions in the double-null coordinates. We discussed why the usual Schwarzschild spherically system coordinates are not a good choice to investigate the critical collapse and showed how the double-null coordinates can improve the accuracy and reduce the performance time in numerical simulations. To examine our solver, we made some tests to check the convergency and accuracy. Results showed that our solver can give out third order convergency. When the apparent horizon begins to form, the grids are focused on the region near apparent horizon automatically, which make us obtain a high precision for the position of initial apparent horizon with a few grids.

Then we used the new algorithm to study the multiple critical gravitational collapse of charged massless scalar in an asymptotically flat space-time with a reflecting mirror. We found that the mass and  charge contained in the initial horizon both approach to zero for the supercritical solutions and the initial black hole is nearlly neutral  obeying $Q_h\ll M_h$ for each critical solution.  In the case that $\epsilon\rightarrow\epsilon_{n0}^-$,  $Q_h$ and $M_h$ both have gaps. We studied the scaling behavior and computed the critical exponents for the first two critical solutions when $\epsilon\rightarrow \epsilon^\pm_{n0}$ , respectively. These critical exponents satisfy the scaling laws proposed by \cite{Cai:2015b}. In addition, the exponents $\beta_{n0}^-$ in our case have a different value  compared with the case in the asymptotically AdS space-time.

For the case of $\epsilon\rightarrow\epsilon_{n0}^+$, we found that the periodic structure appears both in the mass scaling relation and charge scaling relation for the first two critical solutions. These mean that the discrete self-similarity appears for all critical solutions when $\epsilon\rightarrow\epsilon_{n0}^+$. However, in the case of $\epsilon\rightarrow\epsilon_{n0}^-$, our numerical solver cannot give out credible information about sub-leading terms. As the periodic term with its period strongly depends on the discrete echoing character of the critical solution, to find such sub-leading corrections can help us understand which type of self-similarity plays a role when $\epsilon\rightarrow\epsilon_{n0}^-$. The critical exponents for $\epsilon\rightarrow\epsilon_{n0}^+$ can be understood well from the Lyapunov's exponents and renormalization group \cite{Koike:1995jm}. The periodic structure in the their sub corrections can also be understood by treating the Choptuik's solution as an eigenvalue problem \cite{Gundlach:1995kd}. However, to understand the gapped critical behaviors in closed system is still poor. It l needs more investigation in future.

%%%%%%%%%%%%%%%%%%%%%%%%%%%%%%%%%%%%%%%%%%%%%%%%
\section*{Acknowledgements}
This work was supported in part by the National Natural Science Foundation of China (No.11375247 and No.11435006 ).

%%%%%%%%%%%%%%%%%%%%%%%%%%%%%%%%%%%%%%%%%%%%%%%%%%%

%%%%%%%%%%%%%%%%%%%%%%%%%%%%%%%%


\begin{thebibliography}{99}
%\cite{Choptuik}
\bibitem{Choptuik}
M. W. Choptuik,
``Universality and Scaling in Gravitational Collapse of a Massless Scalar Field,"
Phys. Rev. Lett. {\bf 70}, 9 (1993).

%\cite{Gundlach:2007gc}
\bibitem{Gundlach:2007gc}
  C.~Gundlach and J.~M.~Martin-Garcia,
  ``Critical phenomena in gravitational collapse,''
  Living Rev.\ Rel.\  {\bf 10}, 5 (2007)
  doi:10.12942/lrr-2007-5
  [arXiv:0711.4620 [gr-qc]];
  %%CITATION = doi:10.12942/lrr-2007-5;%%
  %85 citations counted in INSPIRE as of 28 Nov 2015
   A.~Wang,
  ``Critical phenomena in gravitational collapse: The Studies so far,''
  Braz.\ J.\ Phys.\  {\bf 31}, 188 (2001)
  doi:10.1590/S0103-97332001000200009
  [gr-qc/0104073].
  %%CITATION = doi:10.1590/S0103-97332001000200009;%%
  %27 citations counted in INSPIRE as of 27 janv. 2016


%\cite{Patrick}
\bibitem{Patrick}
Patrick R Brady
``Analytic example of critical behaviour in scalar field collapse,"
Class. Quantum Grav. {\bf 11}  1255-3260 (1994).

%\cite{Zhang:2014dfa}
\bibitem{Zhang:2014dfa}
  X.~Zhang and H.~L¨¹,
  ``Critical Behavior in a Massless Scalar Field Collapse with Self-interaction Potential,''
  Phys.\ Rev.\ D {\bf 91}, no. 4, 044046 (2015)
  doi:10.1103/PhysRevD.91.044046
  [arXiv:1410.8337 [gr-qc]].
  %%CITATION = doi:10.1103/PhysRevD.91.044046;%%
  %5 citations counted in INSPIRE as of 20 Nov 2015

%\cite{Maison:1995cc}
\bibitem{Maison:1995cc}
  D.~Maison,
  ``Nonuniversality of critical behavior in spherically symmetric gravitational collapse,''
  Phys.\ Lett.\ B {\bf 366}, 82 (1996)
  [gr-qc/9504008].
  %%CITATION = doi:10.1016/0370-2693(95)01381-4;%%
  %90 citations counted in INSPIRE as of 09 Dec 2015

%\cite{Gundlach:1995kd}
\bibitem{Gundlach:1995kd}
  C.~Gundlach,
  ``The Choptuik space-time as an eigenvalue problem,''
  Phys.\ Rev.\ Lett.\  {\bf 75}, 3214 (1995)
  doi:10.1103/PhysRevLett.75.3214
  [gr-qc/9507054].
  %%CITATION = doi:10.1103/PhysRevLett.75.3214;%%
  %68 citations counted in INSPIRE as of 28 Nov 2015

%\cite{Koike:1995jm}
\bibitem{Koike:1995jm}
  T.~Koike, T.~Hara and S.~Adachi,
  ``Critical behavior in gravitational collapse of radiation fluid: A Renormalization group (linear perturbation) analysis,''
  Phys.\ Rev.\ Lett.\  {\bf 74}, 5170 (1995)
  doi:10.1103/PhysRevLett.74.5170
  [gr-qc/9503007].
  %%CITATION = doi:10.1103/PhysRevLett.74.5170;%%
  %137 citations counted in INSPIRE as of 28 Nov 2015

%\cite{Maldacena:1997re}
\bibitem{Maldacena:1997re}
  J.~M.~Maldacena,
  ``The Large N limit of superconformal field theories and supergravity,''
  Int.\ J.\ Theor.\ Phys.\  {\bf 38}, 1113 (1999)
  [Adv.\ Theor.\ Math.\ Phys.\  {\bf 2}, 231 (1998)]
  doi:10.1023/A:1026654312961
  [hep-th/9711200].
  %%CITATION = doi:10.1023/A:1026654312961;%%
  %11235 citations counted in INSPIRE as of 28 Nov 2015

%\cite{Witten:1998qj}
\bibitem{Witten:1998qj}
  E.~Witten,
  ``Anti-de Sitter space and holography,''
  Adv.\ Theor.\ Math.\ Phys.\  {\bf 2}, 253 (1998)
  [hep-th/9802150].
  %%CITATION = HEP-TH/9802150;%%
  %7431 citations counted in INSPIRE as of 28 Nov 2015

%\cite{Bizon:2011gg}
\bibitem{Bizon:2011gg}
  P.~Bizon and A.~Rostworowski,
  ``On weakly turbulent instability of anti-de Sitter space,''
  Phys.\ Rev.\ Lett.\  {\bf 107}, 031102 (2011)
  [arXiv:1104.3702 [gr-qc]].
  %%CITATION = ARXIV:1104.3702;%%
  %174 citations counted in INSPIRE as of 12 Oct 2015

%\cite{Maliborski:2012gx}
\bibitem{Maliborski:2012gx}
  M.~Maliborski,
  ``Instability of Flat Space Enclosed in a Cavity,''
  Phys.\ Rev.\ Lett.\  {\bf 109}, 221101 (2012)
  [arXiv:1208.2934 [gr-qc]].
  %%CITATION = ARXIV:1208.2934;%%
  %25 citations counted in INSPIRE as of 08 Nov 2015

%\cite{Olivan:2015fmy}
\bibitem{Olivan:2015fmy}
  D.~S.~Oliv\'{a}n and C.~F.~Sopuerta,
  ``New features of gravitational collapse in Anti-de Sitter spacetimes,''
  arXiv:1511.04344 [gr-qc].
  %%CITATION = ARXIV:1511.04344;%%

%\cite{Evnin:2015gma}
\bibitem{Evnin:2015gma}
  O.~Evnin and C.~Krishnan,
  ``A Hidden Symmetry of AdS Resonances,''
  Phys.\ Rev.\ D {\bf 91}, no. 12, 126010 (2015)
  [arXiv:1502.03749 [hep-th]].
  %%CITATION = ARXIV:1502.03749;%%
  %5 citations counted in INSPIRE as of 12 Oct 2015

%\cite{Mcode}
\bibitem{Mcode}
The main file and some examples for how to use this solver can be downloaded from this link:\\
http://gc.itp.ac.cn/documents/24005/2ea053f8-8290-48de-9af8-179bdf77c5dd

%\cite{Goldwirth}
\bibitem{Goldwirth}
D. S. Goldwirth, and T. Piran,
``Gravitational collapse of massless scalar field and cosmic censorship,"
Phys. Rev. D {\bf 36}, 3575 (1987).

%\cite{Gundlach}
\bibitem{Gundlach}
C. Gundlach, R. H. Price, and J. Pullin,
``Late-time behavior of stellar collapse and explosions. II. Nonlinear evolution,"
Phys. Rev. D {\bf 49}, 890 (1994).

%\cite{Garfinkle}
\bibitem{Garfinkle}
D. Garfinkle,
``Choptuik scaling in null coordinates,"
Phys. Rev. D {\bf 51}, 5558 (1995).

%\cite{Hod:1996ar}
\bibitem{Hod:1996ar}
  S.~Hod and T.~Piran,
  ``Critical behavior and universality in gravitational collapse of a charged scalar field,''
  Phys.\ Rev.\ D {\bf 55}, 3485 (1997)
  [gr-qc/9606093].
  %%CITATION = GR-QC/9606093;%%
  %40 citations counted in INSPIRE as of 08 Nov 2015

%\cite{Cai:2015b}
\bibitem{Cai:2015b}
  R.~G.~Cai and R.~Q.~Yang,
  ``Scaling Laws in Gravitational Collapse,''
  arXiv:1512.07095 [gr-qc].
  %%CITATION = ARXIV:1512.07095;%%


%\cite{Gundlach:1996vv}
\bibitem{Gundlach:1996vv}
  C.~Gundlach and J.~M.~Martin-Garcia,
  ``Charge scaling and universality in critical collapse,''
  Phys.\ Rev.\ D {\bf 54}, 7353 (1996)
  doi:10.1103/PhysRevD.54.7353
  [gr-qc/9606072].
  %%CITATION = doi:10.1103/PhysRevD.54.7353;%%
  %25 citations counted in INSPIRE as of 18 Nov 2015

  %\cite{Zhang:2015dwu}
\bibitem{Zhang:2015dwu}
  C.~Y.~Zhang, S.~J.~Zhang, D.~C.~Zou and B.~Wang,
  ``Charged scalar gravitational collapse in de Sitter spacetime,''
  arXiv:1512.06472 [gr-qc].
  %%CITATION = ARXIV:1512.06472;%%
  %1 citations counted in INSPIRE as of 29 Jan 2016


%\cite{Liebling:2012gv}
\bibitem{Liebling:2012gv}
  S.~L.~Liebling,
  ``Nonlinear collapse in the semilinear wave equation in AdS space,''
  Phys.\ Rev.\ D {\bf 87}, no. 8, 081501 (2013)
  [arXiv:1212.6970 [gr-qc]].
  %%CITATION = ARXIV:1212.6970;%%
  %5 citations counted in INSPIRE as of 11 Nov 2015

%\cite{Cai:2015jbs}
\bibitem{Cai:2015jbs}
  R.~G.~Cai, L.~W.~Ji and R.~Q.~Yang,
  ``Collapse of self-interacting scalar field in anti-de Sitter space,''
  arXiv:1511.00868 [gr-qc].
  %%CITATION = ARXIV:1511.00868;%%






%\cite{Hod:1996az}
\bibitem{Hod:1996az}
  S.~Hod and T.~Piran,
  ``Fine structure of Choptuik's mass scaling relation,''
  Phys.\ Rev.\ D {\bf 55}, 440 (1997)
  doi:10.1103/PhysRevD.55.440
  [gr-qc/9606087].
  %%CITATION = doi:10.1103/PhysRevD.55.440;%%
  %47 citations counted in INSPIRE as of 29 Nov 2015

\end{thebibliography}
\end{document}